\documentclass[10pt,aps,prd,amsmath,amssymb,oneside,showpacs,nofootinbib,preprintnumbers]{revtex4}
\usepackage[T1]{fontenc}
\usepackage{mathrsfs}
\usepackage{natbib}

\newcommand{\defeq}{:=}
\newcommand{\im}{\mathrm{i}}  

\newcommand{\xd}{\mathrm{d}}
\newcommand{\xD}{\mathcal{D}} 

\newcommand{\be}{\begin{equation}}
\newcommand{\ee}{\end{equation}}
\newcommand{\bea}{\begin{eqnarray}}
\newcommand{\eea}{\end{eqnarray}}

\begin{document}

\title{General boundary quantum field theory in de Sitter spacetime}
\author{Daniele Colosi}\email{colosi@matmor.unam.mx}
\affiliation{Instituto de Matem\'aticas,\\ Universidad Nacional Aut\'onoma de M\'exico,\\ Campus Morelia, C.P.~58190, Morelia, Michoac\'an, Mexico}
\date{\today}
\preprint{UNAM-IM-MOR-2010-2}

\begin{abstract}
We quantize a massive scalar field in de Sitter spacetime and derive the $S$-matrix for the general interacting theory. Using the general boundary formulation of quantum field theory, we also propose a new type of $S$-matrix derived from the asymptotic limit of the amplitude associated with a spacetime region bounded by one connected and timelike hypersurface. Based on previous works in Minkowski spacetime, we call this region the hypercylinder region. We show how the new $S$-matrix coincides with the usual one by constructing an isomorphism between the space of temporal asymptotic states of the traditional setting and the space of spatial asymptotic states defined on the asymptotic hypercylinder.
\end{abstract}

\maketitle

\section{Introduction}

The present paper is devoted to the study of the quantization of a real massive scalar field in de Sitter spacetime and to derive the S-matrix for a general interaction within the framework of the general boundary formulation (GBF) of quantum field theory (QFT). The GBF has recently emerged as a new powerful tool to describe the dynamics of quantum fields. The key idea on which the GBF is based resides on a generalization of the notion of amplitudes. To fully appreciate the novelties introduced by the GBF, it would be useful to quickly review how evolution and transition amplitudes are usually described.
In standard Minkowski-based quantum field theory, transition amplitudes are expressed as the scalar product between a state defined on an initial equal-time hyperplane, say $\xi$, evolved up to a final equal-time hyperplane by the evolution operator $U(t_f,t_i)$, and a state defined on the final hyperplane, $\eta$, namely in Dirac notation
\be
\langle \eta, t_f | U(t_f,t_i) |\xi, t_i \rangle.
\label{eq:standardTA}
\ee
The above expression is interpreted as the probability amplitude for the transition from the state $\xi$ to the state $\eta$ in the time interval ($t_f - t_i$). This represents the standard picture of dynamics understood as evolution from an initial state to a final one. The physical process described by (\ref{eq:standardTA}) involves the non compact region of Minkowski spacetime bounded by the two disconnected spacelike hyperplanes defined by the times $t_i$ and $t_f$ respectively. 
Now, if we are interested in the study of process involving a spacetime region naturally bounded by, say, one connected hypersurface containing timelike parts, the application of the above formalism turns out to be problematic: Indeed it is the standard notion of evolution that appears to be questionable since in this case the distinction between initial and final states is missing due to the connectedness of the boundary and a more general notion of evolution is needed.
Other problems may appear in the usual S-matrix technique to calculate probabilities for scattering processes, in which one usually assumes that the interaction vanishes at asymptotic times, so that the initial and final state, now defined on temporal asymptotic hyperplanes, belong to the state space of the corresponding free theory. 
But how can we define the S-matrix if the interaction never vanishes or if for some reasons no temporal asymptotic states exist (as will be the case in anti-de Sitter spacetime)? 
Moreover, from an experimental point of view, infinitely distant regions are inaccessible, and real experiments always take place in finite regions of spacetime. This observation opens the question of the implementability of a fully local description of the dynamics of quantum fields, namely a description involving only compact spacetime regions\footnote{In \citep{ColRov:localpar} a notion of local particle states, understood as quanta of a local field operator, was introduced and its relation with the standard Fock particle states analized.}. While this is difficult or even incompatible with the standard techniques of QFT\footnote{By standard QFT we mean the Hilbert-space approach of QFT, in which the central object is the vector space of states given by a complex Hilbert space, with observables represented by self-adjoint operators acting on it. We do not refer here to the algebraic approach to QFT.}, the GBF offers the appropriate framework for such description.
More serious problems emerge in the study of quantum fields in curved spacetimes, apart from the one just mentioned relative to anti-de Sitter space. The class of problems we are referring to originates form the absence (in general) of symmetries of the spacetime metric preventing the availability of a privileged criterion to select a specific vacuum state of the field, which leads consequently to many inequivalent quantum theories. Furthermore, in this case, evolution is understood between arbitrary spacelike Cauchy surfaces that provide a foliation of the spacetime and in general results to be non-unitary \citep{BiDa:qfcs}.
The description of physical processes in a background-independent quantum context, as would be the case in quantum gravity, will entail even more radical problems requiring drastic departure from both technical and conceptual aspects of standard QFT\footnote{Quantum gravity was indeed among the main  motivations for the GBF. All the interest in a general boundary approach to quantum gravity resides in the possibility that the GBF can handle some of the major conceptual problems posed by any quantum theory not defined on a fixed background metric, such as the problem of time and the problem of locality \citep{Oe:bqgrav}. By the problem of locality we mean the separation of the system of interest from the rest of the universe. While this is possible within quantum field theory, due to causality and the cluster decomposition principle, in the absence of background metric distant systems can not be separated and treated independently.}. 

The GBF not only brings a new viewpoint on QFT but may also solve some of the problems mentioned above. Indeed within the GBF a consistent description is implemented for physical processes taking place in \textsl{arbitrary} spacetime regions that are not consequently restricted to those involved in equation (\ref{eq:standardTA}), namely regions defined by a time interval. The major novelty is then represented by considering regions with compact boundary, i.e. regions with a boundary hypersurface having both spacelike and timelike components. On the other hand, standard QFT can be recovered from the GBF by specializing the boundary hypersurface to the disjoint union of two Cauchy surfaces. In that sense the GBF represents an extension of the quantum theory. The basic ingridients of this new formulation are inspired by topological quantum field theory: state spaces are associated with the boundary of any spacetime region, and amplitudes are associated with the region and are given by maps from these state spaces to the complex numbers. These structures are required to satisfy a set of axioms assuring their consistency. Section \ref{sec:qt} is devoted to the elucidation of these state spaces and amplitudes, which are introduced using the Schr\"odinger representation for the quantum states of the field, combined with the Feynman path integral quantization prescription\footnote{It is important to mention that the GBF is a general framework to formulate quantum theories and is not bound to any particular theory. Nor to any specific quantization scheme consequently. However, a Schr\"odinger-Feynman quantization turns out to be useful and to work, at least at a formal level, in the specific situations considered so far. We mention the existence of new and more rigorous quantization scheme proposed by Oeckl in \citep{Oe:newGBF}.}. Finally, a consistent physical interpretation can be given to such amplitudes and an appropriate notion of probability can be extracted from them \citep{Oe:GBQFT,Oe:KGtl}. It is important to emphasize that the GBF implements a manifestly local description of the quantum theory: Indeed the computation of the amplitudes takes into account only the states defined on the boundary and the dynamics, compatible with the specified boundary states, taking place in the spacetime region of interest.  

In a series of papers the GBF has been applied to study interacting scalar fields in Minkowski \citep{CoOe:spsmatrix,CoOe:smatrixgbf} and Euclidean spacetime \citep{CoOe:smatrix2d}. On the one hand the usual results of standard QFT have been recovered by considering time interval regions. On the other hand dynamics taking place in a new type of spacetime regions, namely regions bounded by a timelike and \textsl{connected} hypersurface, were described and a consistent probability interpretation was implemented. In Minkowski spacetime the region considered was a timelike hypercylinder, i.e. a ball in space extended over all of time, first introduced in \citep{Oe:KGtl}. In the 2-dimensional Euclidean space of \citep{CoOe:smatrix2d} the region was given by a circle. Notice that the connectedness of the boundary prevents a natural decomposition of the state space associated with it into a state space containing \textsl{in}-states and one containing \textsl{out}-states. Thus, the traditional picture of dynamics, entails by expression (\ref{eq:standardTA}), has to be extended to a more general one. The key point is that the GBF provides a precise mathematical description (at the level of rigor of the path integral) as well as a suitable physical interpretation for processes characterized by these geometries. Furthermore, it has been shown the existence of an isomorphism between the state spaces associated with the boundaries of the hypercylinder in Minkowski space with the state space associated with the time interval region. Due to this isomorphism the interacting asymptotic amplitude defined from the large radius limit of the hypercylinder results to be equivalent to the S-matrix in the standard setting, when both can be defined. Analogue results were obtained for field theory in Euclidean space.

The cases studied so far are restricted to flat-spacetime-based QFT. Here we take a further step by considering the quantum theory of a scalar field in a curved spacetime, specifically in de Sitter spacetime. The de Sitter spacetime presents many interesting features. First, because of its high degree of symmetry, the dynamics of quantized fields propagating in it is exactly solvable and many of the properties of the fields can be studied analytically \citep{BiDa:qfcs}. Moreover, in the inflationary cosmology scenario de Sitter space models an exponentially expanding universe at the initial stage of inflation \citep{Mukhanov:2005sc}. 
De Sitter space has also attracted new interest in connection with the conjecture of the dS/CFT correspondence proposed almost a decade ago by Strominger \citep{Strominger:2001pn}.

In this article we will quantize a real massive minimally coupled scalar field in de Sitter spacetime within the GBF. The main result consists in the derivation of the expression of the S-matrix for the general interacting theory in two different contexts. The first one is represented by the standard settings of QFT, where the field evolves between spacelike Cauchy surfaces. The second one is inspired by the geometry of the hypercylinder in Minkowski space: evolution takes place in a non-compact spacetime region bounded by one timelike connected hypersurface on which the quantum states of the field are defined. In particular this hypersurface is at fixed distance from the temporal axis of de Sitter space, and will be called the hypercylinder in analogy to the Minkowski case. The S-matrix for the hypercylinder region is given by the asymptotic amplitude (at spatial infinity) for the specified quantum states. These two S-matrices will be shown to be equivalent by the existence of an isomorphism between the state spaces associated with the corresponding asymptotic hypersurfaces, hence extending to de Sitter space the results of \citep{CoOe:spsmatrix,CoOe:smatrixgbf,CoOe:smatrix2d}. Part of the results presented here were announced in a previous paper \citep{Co:dS}.

The outline of the paper is as follows. In Sec. \ref{sec:classical} we present the two spacetime regions we will be interested in and solve the Klein-Gordon equation in the different coordinate systems chosen in the two regions. In Sec. \ref{sec:qt} the main structures of the GBF are introduced. In Sec. \ref{sec:free} the free scalar field is quantized in the region with spacelike boundary, namely the time interval region bounded by two equal-time hypersurfaces. We start by expressing the field propagator and then introduce the vacuum state and coherent states. Finally the asymptotic amplitude, interpreted as the $S$-matrix for the free theory is computed. We then treat the interacting theory in Sec. \ref{sec:int} in two steps: First the asymptotic amplitude in the case of an interaction with a source field is derived and subsequently we use functional methods to work out the $S$-matrix for the general interacting theory. In Sec. \ref{sec:hypfree} we quantize the free field in the hyercylinder region, following the treatment of Sec. \ref{sec:free}: after introducing the main structures we obtain the expression of the asymptotic amplitude for coherent states defined on the hypercylinder. Sec. \ref{sec:hypint} deals with the interacting theory and the asymptotic amplitude is derived following the same steps of Sec. \ref{sec:int}. In Sec. \ref{sec:iso} we show the existence of an isomorphism between the asymptotic Hilbert spaces associated with the boundaries of the two geometries in question and prove the equivalence of the asymptotic amplitudes under such isomorphism. Sec. \ref{sec:con} contains our conclusions and an outlook. A few technical details are collected in appendix \ref{sec:apA} and appendix \ref{sec:apB}.

\section{Classical theory}
\label{sec:classical}

We begin in this section by studying the classical theory of a real massive minimally coupled scalar field in de Sitter spacetime. The action in a spacetime region $M$ is given by
\be
S_M(\phi) = \frac{1}{2} \int_M \xd^4 x \sqrt{-g}\left( g^{\mu \nu} \partial_{\mu} \phi \, \partial_{\nu} \phi - m^2 \phi^2 \right),
\label{eq:action0}
\ee
where we use the notation $\partial_{\mu}= \partial / \partial x^{\mu}$, and $g \equiv \det g_{\mu \nu}$ denotes the determinant of the de Sitter metric. Via an integration by parts, the action of a classical solution $\phi_{cl}$ of the equation of motion obtained from (\ref{eq:action0}) reduces to a boundary term,
\be
S_{M}(\phi_{cl}) = \frac{1}{2} \int_{\partial M} \xd ^3 s \sqrt{g^{(3)}} \, \left( \phi_{cl} \, \partial_n \phi_{cl} \right),
\label{eq:actcl}
\ee
where $s$ indicates generic three dimensional coordinates on the boundary $\partial M$ of the region $M$, $\partial_n$ is the normal outward derivative to $\partial M$, namely $\partial_n = n^{\mu} \partial_{\mu}$ with $n^{\mu}$ the normal to the surface and $g^{(3)}$ is the determinant of the 3-metric induced on $\partial M$. In the following subsections two different regions M will be considered. First we will be interested in a region with spacelike boundaries specified by constant values of the de Sitter time (defined below). Then, the second region we will look at is a region with timelike boundaries determined by constant values of the radial distance from the origin of de Sitter spacetime. Our main goal will be to provide an expression for both the action (\ref{eq:actcl}) and the classical field $\phi_{cl}$ in terms of the boundary field configurations $\varphi$,
\be
\phi(x) \big|_{x \in \partial M} = \varphi (s).
\ee

\subsection{Region with spacelike boundary}

The first geometry we will consider is appropriately described in terms of a coordinate system $(t,\underline{x})$, where the de Sitter metric takes the form
\bea
\xd s^2 = \frac{R^2}{t^2} \left( \xd t^2 - \xd \underline{x}^2  \right),
\label{eq:dSmetric}
\eea
where $t \in (0, \infty)$, $\underline{x} \in \mathbb{R}^3$ and $R$ denotes the inverse of the Hubble constant. This coordinate system covers the half of de Sitter space, the remaining half can be included by extending the domain of the conformal time $t$ to negative values.
We consider a region $M$ of de Sitter spacetime bounded by the disjoint union of two hypersurfaces of constant conformal time $t$, namely $\Sigma_1 = \{ (t, \underline{x}) : t= t_1 \}$ and $\Sigma_2 = \{ (t, \underline{x}) : t= t_2 \}$, with $t_1 < t_2$. We denote this spacetime region, $M= [t_1, t_2] \times \mathbb{R}^3$, simply by $[t_1, t_2]$. 

The action of a real massive minimally coupled scalar field $\phi$ in this region $[t_1,t_2]$ is given by
\bea
S_{[t_1, t_2]}(\phi) = \frac{1}{2} \int_{t_1}^{t_2} \xd t \int_{{\mathbb R}^3} \xd^3 \underline{x} \, \frac{R^2}{t^2} \left((\partial_t \phi)^2 - \sum_i (\partial_{i} \phi)^2 - m^2 \phi^2\right).
\label{eq:action}
\eea
The Klein-Gordon equation satisfied by the field $\phi$ has the form
\be
\left[ \frac{t^2}{R^2} \left(\partial_t^2 - \Delta_{\underline{x}} \right) -\frac{2t}{R^2} \partial_t +m^2 \right]\phi(t, \underline{x}) = 0,
\ee
where $\Delta_{\underline{x}}$ is the Laplacian operator in the coordinates $\underline{x}$. This equation can be solved by the method of separation of variables and the general solution can by written as \cite{Schomblond:1976xc}
\be
\phi(t, \underline{x}) =\int \frac{\xd ^3 \underline{k}}{(2 \pi)^{3/2}} \left( v_k(t) \, e^{\im \underline{k} \cdot \underline{x}} +  \overline{v_k}(t) \, e^{-\im \underline{k} \cdot \underline{x}} \right),
\label{eq:classsol}
\ee
where 
\be
v_k(t)= t^{3/2} \left(c_{1}(k) J_{\nu}(kt) + c_{2}(k) Y_{\nu}(kt) \right),
\ee
where $k=|\underline{k}|$, $J_{\nu}(z)$ and $Y_{\nu}(z)$  are the Bessel functions of the first and second kind respectively, with index $\nu = \sqrt{\frac{9}{4} - (m R)^2}$, and $c_{1}(k)$ and $c_{2}(k)$ are two coefficients. In order for the classical solution (\ref{eq:classsol}) to be bounded in the spacetime region $M$ the components of the 3-vector $\underline{k}$ have to be real. Consequently the modulus $k$ is a non negative quantity, $k \geq 0$.

It will be convenient to express the classical solution (\ref{eq:classsol}) in a different form,
\be
\phi(t, \underline{x}) = \left( J_{\nu}(k t) \, \varphi_J \right)(\underline{x}) + \left( Y_{\nu}(k t) \, \varphi_Y \right)(\underline{x}).
\ee
In this expression the Bessel functions $J_{\nu}$ and $Y_{\nu}$ represent operators acting on the field configurations $\varphi_J$ and $\varphi_Y$ respectively. The relation between these field configurations and the boundary field configurations, indicated by $\varphi_1$ and $\varphi_2$ on the hypersurfaces $\Sigma_1$ and $\Sigma_2$ respectively, namely
\be
\varphi_1(\underline{x}) \defeq \phi(t_1,\underline{x}) \qquad \hbox{and} \qquad \varphi_2(\underline{x}) \defeq \phi(t_2,\underline{x}),
\ee
is given by the matrix operator equation
\be
 \begin{pmatrix}\varphi_1 \\ \varphi_2\end{pmatrix}
 =\begin{pmatrix}J_{\nu}(k t_1) & Y_{\nu}(k t_1)\\ J_{\nu}(k t_2) & Y_{\nu}(k t_2)\end{pmatrix}
 \begin{pmatrix}\varphi_J \\ \varphi_Y\end{pmatrix}
\ee
Inverting this equation we obtain for the field $\phi$ the following dependence on the boundary field configurations,
\be
\phi(t, \underline{x}) = \left( \frac{\delta_k( t, t_2)}{\delta_k ( t_1, t_2)} \, \varphi_1 \right)(\underline{x}) + \left( \frac{\delta_k ( t_1, t)}{\delta_k ( t_1, t_2)} \, \varphi_2 \right) (\underline{x}),
\label{eq:boundconfig}
\ee
where the quotients have to be understood as operators acting on a Fourier expansion of the boundary configurations $\varphi_1$ and $\varphi_2$, and the operator $\delta_k$ is defined as
\be
\delta_k( z, \hat{z}) \defeq z^{3/2} \, \hat{z}^{3/2} \left[ J_{\nu}(k z) \, Y_{\nu}(k \hat{z}) - Y_{\nu}(k z) \, J_{\nu}(k \hat{z})\right].
\label{eq:delta2}
\ee
The expression (\ref{eq:boundconfig}) allows the evaluation of the action (\ref{eq:action}) for a classical solution of the Klein-Gordon equation in terms of the boundary field configurations $\varphi_1$ and $\varphi_2$. The result is
\be
S_{[t_1, t_2]}(\phi)
=  \frac{1}{2} \int \xd^3 \underline{x} \, 
\begin{pmatrix}\varphi_1 & \varphi_2 \end{pmatrix} W_{[t_1, t_2]}  
\begin{pmatrix} \varphi_1 \\ \varphi_2 \end{pmatrix},
\label{eq:actionbound}
\ee
where the $W_{[t_1, t_2]}$ is a 2x2 matrix with elements $W_{[t_1, t_2]}^{(i,j)}, (i,j=1,2),$ given by
\bea
W_{[t_1, t_2]}^{(1,1)} &=& - \frac{R^2}{t_1^2} \left( \frac{3}{2 t_1} + k \frac{J_{\nu}'(k t_1) \, Y_{\nu}(k t_2) - Y_{\nu}'(k t_1) \, J_{\nu}(k t_2)}{J_{\nu}(k t_1) \, Y_{\nu}(k t_2) - Y_{\nu}(k t_1) \, J_{\nu}(k t_2)}\right), \label{eq:W11} \\
W_{[t_1, t_2]}^{(1,2)} &=& W_{[t_1, t_2]}^{(2,1)} = - \frac{2 R^2}{\pi \delta_k (t_1, t_2)}, \label{eq:W12} \\
W_{[t_1, t_2]}^{(2,2)} &=& \frac{R^2}{t_2^2} \left( \frac{3}{2 t_2} + k \frac{J_{\nu}(k t_1) \, Y_{\nu}'(k t_2) - Y_{\nu}(k t_1) \, J_{\nu}'(k t_2)}{J_{\nu}(k t_1) \, Y_{\nu}(k t_2) - Y_{\nu}(k t_1) \, J_{\nu}(k t_2)}\right),\label{eq:W22}
\eea
where a prime indicates the derivative with respect to the argument. These matrix elements have to be understood as operators acting on the boundary field configurations.

\subsection{Region with timelike boundary}
\label{sec:hyp}

The second geometry we are interested in is conveniently described in terms of spherical coordinates in space, defined by three parameters: $r \in [0, \infty)$, $\theta \in [0, \pi)$ and $\varphi \in [0, 2 \pi)$. The de Sitter metric (\ref{eq:dSmetric}) in this coordinate system takes the form
\be
\xd s^2 = \frac{R^2}{t^2} \left( \xd t^2 - \xd r^2 - r^2 \xd \vartheta^2 -r^2 \sin^2 \vartheta \, \xd \varphi^2 \right).
\label{eq:dSmetric2}
\ee
It will be useful in the following to adopt $\Omega$ as a collective notation for $\theta$ and $\varphi$. 
Two different spacetime regions will be considered: The first region is bounded by one hypersurface of radius $r$, denoted by $\Sigma_{\varrho}= \{(t,r,\Omega) : r =\varrho  \}$. In analogy with the case of Minkowski spacetime, we refer to the hypersurface $\Sigma_{\varrho}$ as the hypercylinder of radius $\varrho$. The second region is the spacetime region in between two hypercylinders of different radii $\Sigma_{\varrho}$ and $\Sigma_{\hat{\varrho}}$. Both these regions have timelike boundary, in contrast to the region $[t_1,t_2]$ of the previous subsection. Moreover, the region enclosed by one hypercylinder has a more exotic property: its boundary is not the disjoint union of two disconnected hypersurfaces, it is completely connected.

The Klein-Gordon equation in the metric (\ref{eq:dSmetric2}) reads
\be
\left( \frac{t^2}{R^2} \left[ \partial_t^2 - \Delta_r - \Delta_{\Omega}\right] -\frac{2 t}{R^2} \partial_t + m^2 \right)\phi(t,r, \Omega) = 0,
\label{eq:KGhyp}
\ee
where
\be
\Delta_r= \frac{1}{r^2} \, \partial_r (r^2 \, \partial_r), \qquad \mbox{and} \qquad
\Delta_{\Omega} =\frac{1}{r^2 \, \sin \vartheta} \, \partial_{\vartheta} (\sin \vartheta \, \partial_{\vartheta}) +\frac{1}{r^2 \, \sin^2 \vartheta} \, \partial_{\varphi}^2.
\ee
The bounded solutions of (\ref{eq:KGhyp}) in the region bounded by one or two hypercylinders can be expanded as follows
\be
\phi(t,r,\Omega) = \int_{-\infty}^{\infty} \xd k \, \sum_{l=0}^{\infty} \sum_{m=-l}^l \left( a_{k,l,m} \, u_{k,l,m}(t,r,\Omega) + c.c. \right),
\label{eq:classsolhyp}
\ee
where $a_{k,l,m}$ are coefficients and with $u_{k,l,m}$ we denote the unnormalized modes
\be
u_{k,l,m}(t,r,\Omega) =  t^{3/2} \, {\mathscr H}_{\nu}(k t) \, Y_l^m(\Omega) \left( c_1 (k)j_l(k r) + c_2(k) n_l(k r) \right).
\label{eq:uklm}
\ee
The coefficients $c_1$ and $c_2$ are in general different from those introduced in the previous section. $Y_l^m$ are the spherical harmonics satisfying the equation
\be
\left( \Delta_{\Omega}  Y_l^m \right) (\Omega) = -\frac{l(l+1)}{r^2} Y_l^m(\Omega).
\ee
In the modes (\ref{eq:uklm}), $j_l$ and $n_l$ denote the spherical Bessel functions of the first and second kind respectively, solutions of the equation
\be
\left( \Delta_r j_l\right) (k r) = \left( \frac{l(l+1)}{r^2} -k^2 \right) j_l(kr),
\label{eq:radialKG}
\ee
and the same equation is satisfied by $n_l$. 
Notice that for the spacetime region enclosed by the hypercylinder, where the origin ($r=0$) is part of the region, the coefficient $c_2$ in (\ref{eq:uklm}) will be zero and the radial component of the modes $u_{k,l,m}$ will reduce to the spherical Bessel function of the fist kind, $j_l$. The reason lies in the singular character of $n_l$ in the origin, whereas $j_l$ remains finite \cite{AbSt:handbook}. On the other hand, the spacetime region bounded by two hypercylinders does not contain the origin, and both $j_l$ and $n_l$ will appear in the modes $u_{k,l,m}$. 

Finally, the function ${\mathscr H}_{\nu}$ in (\ref{eq:uklm}) is proportional to the Bessel functions of the third kind of order $\nu$, $H_{\nu}$, as called Hankel function\footnote{Working with ${\mathscr H}_{\nu}$ instead of $H_{\nu}$ turns out to be more convenient due to the following property,
\be
{\mathscr H}_{\nu} ( - kt)= \overline{{\mathscr H}_{\nu}} (kt).
\label{eq:modHankel}
\ee
This relation follows from the analytic continuation of the Hankel function \cite{Wat:bessel} and will be used in many occasions in the rest of paper.},
\be
{\mathscr H}_{\nu} (kt)= e^{\im \nu \pi/2} H_{\nu}(kt).
\ee 
In order for the Hankel function to be bounded $k$ must be real.

Consider the region with the hypercylinder of radius $\varrho$ as boundary. In the following we will refer to this region simply by $\varrho$.\footnote{The symbol $\varrho$ should not be confused with the symbol $\rho$ denoting the amplitude in the subsequent sections.} The classical solution of (\ref{eq:KGhyp}) matching the boundary field configuration $\varphi$ on the hypercylinder, i.e. for $r= \varrho$, can be written as
\be
\phi(t,r,\Omega) = \left( \frac{j_l(k r)}{j_l(r \varrho)} \, \varphi \right) (t, \Omega),
\label{eq:hypbound}
\ee
where the quotient of spherical Bessel functions as to be understood as an operator. The action (\ref{eq:actcl}) associated with the region $\varrho$ takes the form
\be
S_{\varrho} (\phi)=  -\frac{1}{2} \int \xd t \, \xd \Omega \, \frac{R^2}{t^2} \, \varrho^2 \, \varphi(t, \Omega)  \left( k \,\frac{j_l'(k \varrho)}{j_l(k \varrho)} \, \varphi \right) (t, \Omega),
\label{eq:actbound2}
\ee
where the prime indicates the derivative with respect to the argument.

We now turn to the spacetime region bounded by two hypercylinders of different radii, $\varrho_1$ and $\varrho_2$, indicated by $[\varrho_1, \varrho_2]$.
Let $\varphi_1$ and $\varphi_2$ denote the boundary field configurations on $r=\varrho_1$ and $r=\varrho_2$ respectively. The classical solution of the Klein-Gordon equation reducing to these field configurations on the boundary of $[\varrho_1, \varrho_2]$ is relate to $\varphi_1$ and $\varphi_2$ via
\be
\phi(t,r,\Omega) = \left( \frac{\Delta_k(r, \varrho_2)}{\Delta_k(\varrho_1, \varrho_2)} \varphi_1\right) (t, \Omega) +
\left( \frac{\Delta_k(\varrho_1, r)}{\Delta_k(\varrho_1, \varrho_2)} \varphi_2 \right) (t, \Omega),
\label{eq:boundconfighyp}
\ee
where the function $\Delta_k$ is to be understood as an operator defined by
\be
\Delta_k(\varrho_1, \varrho_2) = j_l(k \varrho_1)n_l(k \varrho_2) - n_l(k \varrho_1)j_l(k \varrho_2).
\ee
The action (\ref{eq:actcl}) of the field (\ref{eq:boundconfighyp}) is then
\be
S_{[\varrho_1, \varrho_2]}(\phi) =  \frac{1}{2} \int \xd t \, \xd \Omega \, 
\begin{pmatrix}\varphi_1 & \varphi_2 \end{pmatrix} {\cal W}_{[\varrho_1, \varrho_2]}  
\begin{pmatrix} \varphi_1 \\ \varphi_2 \end{pmatrix},
\label{eq:actbound3}
\ee
where the ${\cal W}_{[\varrho_1, \varrho_2]}$ is a 2x2 matrix with elements ${\cal W}_{[\varrho_1, \varrho_2]} ^{(i,j)}, (i,j=1,2),$ given by
\bea
{\cal W}_{[\varrho_1, \varrho_2]}^{(1,1)} &=&  \frac{R^2}{t^2} \, \varrho_1^2 \,  \frac{k \,  \sigma_k(\varrho_2, \varrho_1)}{\Delta_k(\varrho_1, \varrho_2)},
\label{eq:Whyp1}\\
{\cal W}_{[\varrho_1, \varrho_2]}^{(1,2)} &=& {\cal W}_{[\varrho_1, \varrho_2]}^{(2,1)} =  - \frac{ R^2}{t^2} \frac{1}{ k \, \Delta_k (\varrho_1, \varrho_2)},
\label{eq:Whyp2}\\
{\cal W}_{[\varrho_1, \varrho_2]}^{(2,2)} &=&  \frac{R^2}{t^2} \, \varrho_2^2 \,  \frac{k \, \sigma_k(\varrho_1, \varrho_2)}{\Delta_k(\varrho_1, \varrho_2)}.
\label{eq:Whyp3}
\eea
The function $\sigma_k$ is to be understood as an operator defined as
\be
\sigma_k(\varrho_1, \varrho_2)= j_l(k \varrho_1)n_l'(k \varrho_2) - n_l(k \varrho_1)j_l'(k \varrho_2).
\ee
The expressions of the action of a classical solution of the equation of motion in terms of the boundary field configurations for the different spacetime regions considered, (\ref{eq:actionbound}), (\ref{eq:actbound2}) and (\ref{eq:actbound3}), will be an important ingredient for the computation of the quantum field propagator, as will be clear in the next section.

\section{Quantum theory}
\label{sec:qt}

According to the axioms of the GBF, a Hilbert space ${\cal H}_{\Sigma}$ of states is associated with each hypersurface $\Sigma$. The quantum states in this Hilbert space are described in the Schr\"odinger representation, namely quantum states are wave functionals on the space of field configurations $K_{\Sigma}$. The inner product in ${\cal H}_{\Sigma}$ is defined through an integral over field configurations,
\be
\langle \psi_{\Sigma} | \psi_{\Sigma}' \rangle \defeq \int_{K_{\Sigma}} \xD \varphi \, \overline{\psi_{\Sigma}(\varphi)} \, \psi_{\Sigma}(\varphi).
\ee 
The evolution of a quantum state $\psi_{\Sigma} \in {\cal H}_{\Sigma}$ to a quantum state $\psi_{\hat{\Sigma}} \in {\cal H}_{\hat{\Sigma}}$ is given in terms of the field propagator $Z_{[\Sigma, \hat{\Sigma}]}$ associated with the spacetime region bounded by the two hypersurfaces $\Sigma$ and $\hat{\Sigma}$,
\be
\psi_{\hat{\Sigma}} (\hat{\varphi}) = \int_{K_{\Sigma}} \xD \varphi \, \psi_{\Sigma}(\varphi) \, Z_{[\Sigma, \hat{\Sigma}]} (\varphi, \hat{\varphi}).
\ee
The field propagator $Z_{[\Sigma, \hat{\Sigma}]}$ encodes the evolution from the field configuration $\varphi$ on the hypersurface $\Sigma$ to the field configuration $\hat{\varphi}$ on the hypersurface $\hat{\Sigma}$. It is defined by the Feynman path integral as
\be
Z_{[\Sigma, \hat{\Sigma}]} (\varphi, \hat{\varphi}) = \int_{\phi|_{\Sigma} = \varphi, \, \phi|_{\hat{\Sigma}} = \hat{\varphi}} \xD \phi \, e^{\im S_{[\Sigma, \hat{\Sigma}]}(\phi)},
\label{eq:prop}
\ee
the integral is extended over all field configurations $\phi$ in the spacetime region bounded by the two hypersurfaces $\Sigma$ and $\hat{\Sigma}$ that reduce to $\varphi$ on $\Sigma$ and to $\hat{\varphi}$ on $\hat{\Sigma}$, and $S_{[\Sigma, \hat{\Sigma}]}$ is the action integral over this spacetime region. The path integral (\ref{eq:prop}) can be formally evaluated by shifting the integration variable by a classical solution $\phi_{cl}$ of the equation of motion derived from $S_{[\Sigma, \hat{\Sigma}]}$, matching the boundary configurations $\varphi$ and $\hat{\varphi}$ on the boundaries of the region. Explicitly,
\be
Z_{[\Sigma, \hat{\Sigma}]} (\varphi, \hat{\varphi}) = \int_{\phi|_{\Sigma} = \varphi, \, \phi|_{\hat{\Sigma}} = \hat{\varphi}} \xD \phi \, e^{\im S_{[\Sigma, \hat{\Sigma}]}(\phi)}
= \int_{\phi|_{\Sigma} = \phi|_{\hat{\Sigma}} = 0} \xD \phi \, e^{\im S_{[\Sigma, \hat{\Sigma}]}(\phi+ \phi_{cl})}
= N_{[\Sigma, \hat{\Sigma}]} e^{\im S_{[\Sigma, \hat{\Sigma}]}( \phi_{cl})},
\label{eq:propcl}
\ee
where the normalization faction is formally given by
\be
N_{[\Sigma, \hat{\Sigma}]} = \int_{\phi|_{\Sigma} = \phi|_{\hat{\Sigma}} = 0} \xD \phi \, e^{\im S_{[\Sigma, \hat{\Sigma}]}(\phi)}.
\ee
Finally, an amplitude $\rho_{[\Sigma, \hat{\Sigma}]}$ is associated with the spacetime region $[\Sigma, \hat{\Sigma}]$ and a state $\psi_{\Sigma} \otimes \overline{\psi_{\hat{\Sigma}}}$ in the Hilbert space associated with the boundary $\partial [\Sigma, \hat{\Sigma}]$, ${\cal H}_{\partial [\Sigma, \hat{\Sigma}]} = {\cal H}_{\Sigma} \otimes {\cal H}_{\hat{\Sigma}}^*$. The amplitude $\rho_{[\Sigma, \hat{\Sigma}]} : {\cal H}_{\partial [\Sigma, \hat{\Sigma}]} \rightarrow {\mathbb C}$ is defined as
\be
\rho_{[\Sigma, \hat{\Sigma}]} (\psi_{\Sigma} \otimes \overline{\psi_{\hat{\Sigma}}})= \int \xD \varphi \, \xD \hat{\varphi} \, \psi_{\Sigma}(\varphi) \, \overline{\psi_{\hat{\Sigma}}(\hat{\varphi})} \, Z_{[\Sigma, \hat{\Sigma}]} (\varphi, \hat{\varphi}).
\label{eq:ampl}
\ee
In the following sections we will apply this formulation to describe the quantum dynamics of a scalar field in de Sitter space. In particular the expressions of the field propagator and amplitudes of quantum states will be explicitly worked out in three cases: we will start with the free theory defined by the free action (\ref{eq:action}), then the interaction with an external source field will be considered, and finally we will use functional derivatives techniques to treat the case of a general interaction.

\section{Region with spacelike boundary - Free theory}
\label{sec:free}

\subsection{Field propagator}
\label{sec:freefieldpropagator}

We evaluate the field propagator associated with the spacetime region $[t_1, t_2]$ of Section \ref{sec:classical}. Substituting the expression (\ref{eq:actionbound}) of the classical action of the scalar field in terms of the boundary field configurations $\varphi_1$ and $\varphi_2$ in (\ref{eq:propcl}) leads to 
\be
Z_{[t_1, t_2],0} (\varphi_1, \varphi_2) = N_{[t_1, t_2],0} \, \exp \left( \frac{\im}{2} \int \xd^3 \underline{x} \, 
\begin{pmatrix}\varphi_1 & \varphi_2 \end{pmatrix} W_{[t_1, t_2]}  
\begin{pmatrix} \varphi_1 \\ \varphi_2 \end{pmatrix}\right),
\label{eq:propt}
\ee
where the normalization factor is determined by the gluing properties of the field propagator (an additional subscript 0 has been written in the field propagator and the normalization factor in order to indicate that these quantities refer to the free theory).
The proof of the consistency of the definition (\ref{eq:propt}) is provided by the composition rule satisfied by the field propagator: The evolution in the variable $t$ from $t_1$ to $t_2$ and subsequently from $t_2$ to $t_3$ must equal the direct evolution from $t_1$ to $t_3$. This composed evolution can be expressed in terms of the following equation,
\be
Z_{[t_1, t_3],0} (\varphi_1, \varphi_3)  = \int \xD \varphi_2 \, Z_{[t_1, t_2],0} (\varphi_1, \varphi_2)  \, Z_{[t_2, t_3],0} (\varphi_2, \varphi_3). 
\label{eq:compprop}
\ee 
The propagator (\ref{eq:propt}) satisfies such composition if the normalization factors corresponding to the three regions are related by
\be
N_{[t_1, t_3],0} = N_{[t_1, t_2],0} \, N_{[t_2, t_3],0} \int \xD \varphi_2 \exp \left( \frac{\im}{2} \int \xd ^3 \underline{x} \, \varphi_2(\underline{x})  \left( \frac{2 R^2}{ \pi} \frac{\delta_k(t_1,t_3)}{\delta_k(t_2,t_3)\delta_k(t_2,t_3)} \varphi_2 \right) (\underline{x})\right).
\ee
The solution turns out to be
\be
N_{[t_1, t_2],0} 
= \det \left( - \frac{\im R^2}{ \pi^2 \delta_k(t_1,t_2)} \right)^{1/2}.
\label{eq:normft}
\ee
We are now in the position to check the unitarity of the quantum evolution implemented by the field propagator. Indeed, in this context the unitarity is translated in the following condition \cite{Oe:GBQFT,Oe:KGtl},
\be
\int \xD \varphi_2 \, \overline{Z_{[t_1, t_2]} (\varphi_1, \varphi_2)} \, Z_{[t_1, t_2]} (\varphi_1', \varphi_2) = \delta(\varphi_1 - \varphi_1').
\label{eq:unitarity}
\ee
For the free field propagator (\ref{eq:propt}) we have
\begin{align}
& \int \xD \varphi_2 \, \overline{Z_{[t_1, t_2],0} (\varphi_1, \varphi_2)} \, Z_{[t_1, t_2],0} (\varphi_1', \varphi_2) \nonumber\\
& =  |N_{[t_1, t_2],0}|^2 \int \xD \varphi_2 \exp \left( \im \int \xd ^3 \underline{x} \, \varphi_2 \, \frac{2 R^2}{ \pi \delta_k(t_1,t_2)}(\varphi_1- \varphi_1') \right) \nonumber\\
& \times \exp \left( \im \int \xd ^3 \underline{x} \, (\varphi_1 + \varphi_1')
\left[ 
- \frac{R^2}{t_1^2} \left( \frac{3}{2 t_1} + k \frac{J_{\nu}'(k t_1) \, Y_{\nu}(k t_2) - Y_{\nu}'(k t_1) \, J_{\nu}(k t_2)}{J_{\nu}(k t_1) \, Y_{\nu}(k t_2) - Y_{\nu}(k t_1) \, J_{\nu}(k t_2)}\right)
\right] 
(\varphi_1 - \varphi_1') \right), \nonumber\\
&=|N_{[t_1, t_2],0}|^2 \, \det \left( - \frac{2 \im R^2}{ \pi^2 \delta_k(t_1,t_2)} \right)^{1/2} \, \delta(\varphi_1 - \varphi_1').
\end{align}
Using (\ref{eq:normft}), the product of the first two terms on the right-hand side above equals 1, and consequently the condition (\ref{eq:unitarity}) is verified. Therefore we conclude that the field propagator (\ref{eq:propt}) implements a unitary quantum evolution in the variable $t$ for the free scalar field in de Sitter space. This was shown in \cite{CoOe:unit} in a general setting, but we nevertheless present the detailed steps here and verify them independently.

A well known property of the de Sitter space is that in the limit in which the curvature goes to zero de Sitter space tends to Minkowski space. Recalling that the Ricci scalar is proportional to $R^{-2}$ \cite{BiDa:qfcs}, we can recover the Minkowski metric from the de Sitter metric (\ref{eq:dSmetric}) by considering the limits $R \rightarrow \infty$ and $t \rightarrow \infty$ in such a way that $R/t \rightarrow 1$,
\be
\lim_{\stackrel{R \rightarrow \infty}{t \rightarrow \infty}} \frac{R}{t} = 1.
\label{eq:dStoM}
\ee
We now show that the free field propagator (\ref{eq:propt}) reduces to the free field propagator in Minkowski space evaluated in \cite{CoOe:smatrixgbf}. The Bessel functions of the first and second kind have the following asymptotic expansions for large values of their argument \cite{AbSt:handbook},
\bea
J_{\nu}(x) &=& \sqrt{\frac{2}{\pi x}} \left( \cos \left(x- \nu \frac{\pi}{2} - \frac{\pi}{4} \right) + o(x^{-1})\right), \label{eq:Jasympt}\\
Y_{\nu}(x) &=& \sqrt{\frac{2}{\pi x}} \left( \sin \left(x- \nu \frac{\pi}{2} - \frac{\pi}{4} \right) + o(x^{-1})\right). \label{eq:Yasympt}
\eea
Taking the limits defined by (\ref{eq:dStoM}) and using the expansions (\ref{eq:Jasympt}) and (\ref{eq:Yasympt}), the matrix elements $W_{[t_1,t_2]}^{(i,j)}$ (\ref{eq:W11},\ref{eq:W12},\ref{eq:W22}) reduce to
\bea
W_{[t_1, t_2]}^{(1,1)} &=& k \, \frac{\cos k(t_2 -t_1)}{\sin k(t_2 -t_1)}, \\
W_{[t_1, t_2]}^{(1,2)} &=& W_{[t_1, t_2]}^{(2,1)} = - \frac{k}{\sin k(t_2 -t_1)}, \\
W_{[t_1, t_2]}^{(2,2)} &=& k \, \frac{\cos k(t_2 -t_1)}{\sin k(t_2 -t_1)}.
\eea
These matrix elements equal those derived in Minkowski space (see \cite{Oe:timelike,Oe:KGtl,CoOe:smatrixgbf}) providing the identification $k = \omega$ holds, where $\omega$ is the operator $\sqrt{- \Delta_{\underline{x}} +m^2}$. Moreover, it is easy to show that the normalization factor (\ref{eq:normft}) reduces to the one in flat spacetime. Therefore we conclude that the free field propagator in de Sitter space (\ref{eq:propt}) coincides with the one in Minkowski space in the limit defined by (\ref{eq:dStoM}).

\subsection{Vacuum state}
\label{sec:vacuum}

We compute the vacuum state on the hypersurface of constant conformal time $t$. The starting point is the following Gaussian ansatz for the vacuum wave function,
\be
\psi_{t,0}(\varphi)=C_{t}  \exp\left(-\frac{1}{2}\int \xd ^3 \underline{x} \, \varphi(\underline{x})(A_{t} \varphi)(\underline{x})\right) ,
\ee
where $C_{t}$ is a normalization factor and $A_{t}$ denotes a family of operators indexed by ${t}$. The explicit form of the operator $A_{t}$ has been given in \cite{Co:vacuum}, and the form of the vacuum wave function results to be 
\be
 \psi_{t,0}(\varphi) = C_{t}  \exp\left(\frac{\im}{2}\int \xd^3 \underline{x} \, \frac{R^2}{t^2} \,  \varphi(\underline{x})\left[ k \frac{H_{ \nu}'(k t)}{H_{\nu}(k t)} + \frac{3}{2 t}\right] \varphi(\underline{x})\right).
\label{eq:vac}
\ee 
The requirement that the vacuum state is normalized to 1 fixes the normalization factor $C_{t}$ up to a phase,
\be
|C_{t}|^{-2} = \int \xD \varphi \exp \left( - \frac{1}{2} \int \xd ^3 x \, \varphi(x) \frac{4 R^2}{\pi t^3 |H_{\nu}(k t)|^2} \varphi(x) \right) = \det \left(\frac{2 R^2}{\pi^2 t^3 |H_{\nu}(k t)|^2} \right)^{-1/2}.
\ee
The phase of the normalization factor is fixed by the relation between two vacuum wave functions defined on hypersurfaces of different conformal time. In particular the vacuum wave function (\ref{eq:vac}) satisfies the identity
\be
\psi_{t_2,0}(\varphi_2) = \int \xD \varphi_1 \, \psi_{t_1,0}(\varphi_1) \, Z_{[t_1,t_2],0}(\varphi_1, \varphi_2),
\ee
which implies the following identity for the normalization factors
\bea
C_{t_2} &=& C_{t_1} \, N_{[t_1,t_2],0} \, \int \xD \varphi_1 \exp \left( - \frac{1}{2} \int \xd ^3 x \, \varphi_1 \left[- \frac{2 \im R^2}{\pi \delta_k(t_1,t_2)} \,  \frac{t_2^{3/2} \,H_{\nu}(k t_2)}{ t_1^{3/2} \, H_{\nu}(k t_1)} \right] \varphi_1 \right), \nonumber\\
&=& C_{t_1} \, N_{[t_1,t_2],0} \, \det \left( - \frac{ \im R^2}{\pi^2 \delta_k(t_1,t_2)} \,  \frac{t_2^{3/2} \,H_{\nu}(k t_2)}{ t_1^{3/2} \, H_{\nu}(k t_1)}\right)^{-1/2}.
\eea
Substituting $N_{[t_1,t_2],0}$ with its expression (\ref{eq:normft}), we arrive at
\be
C_{t_2} = C_{t_1} \, \det \left(  \frac{t_2^{3/2} \,H_{\nu}(k t_2)}{ t_1^{3/2} \, H_{\nu}(k t_1)}\right)^{-1/2}.
\ee
This suggest the following solution for the normalization factor $C_t$,
\be
C_t = \det \left(\frac{\sqrt{2} R}{\pi t^{3/2} \, H_{\nu}(k t)} \right)^{1/2}.
\ee
The asymptotic limit, according to the prescription (\ref{eq:dStoM}), of the vacuum wave function (\ref{eq:vac}) coincides (up to a phase factor) with the vacuum wave function defined on equal time hyperplanes in Minkowski space, namely
\be
\psi_{t,0}(\varphi)=  \det \left(\frac{k \, e^{- \im 2 k t } }{\pi } \right)^{1/4} \exp\left(-\frac{1}{2}\int \xd ^3 \underline{x} \, \varphi(\underline{x})(k \varphi)(\underline{x})\right),
\ee 
providing the identification $k= \omega$ holds (see \cite{Oe:timelike,Oe:KGtl,CoOe:smatrixgbf}).

\subsection{Coherent states}

In previous works \cite{CoOe:spsmatrix,CoOe:smatrixgbf,CoOe:smatrix2d,Co:dS} coherent states have been an essential tool for the computation of the asymptotic amplitudes. We follow here the same approach of these works, and define coherent states for the Klein-Gordon field 
in the Schr\"odinger representation on the hypersurface of constant $t$, in terms of a complex function $\eta$ by the expression
\be
\psi_{t, \eta}(\varphi) = K_{t, \eta} \, \exp \left( \int \frac{\xd ^3 \underline{x} \, \xd^3 \underline{k}}{(2 \pi)^3} \, \eta(\underline{k}) \, e^{\im \underline{k}\cdot \underline{x}} \varphi(\underline{x}) \right) \psi_{t,0}(\varphi),
\ee
where the normalization factor $K_{t, \eta}$ is given by
\be
K_{t, \eta} = \exp \left( - \frac{\pi}{8} \int \frac{ \xd^3 \underline{k}}{(2 \pi)^3} \, \frac{t^3}{R^2} |H_{\nu}(k t)|^2 \left( \eta(\underline{k}) \eta(-\underline{k}) + |\eta(\underline{k})|^2\right)  \right).
\ee
The inner product of two coherent states defined by the complex functions $\eta_1$ and $\eta_2$ results to be
\be
\langle \psi_{t, \eta_2} | \psi_{t, \eta_1} \rangle = \exp \left( - \frac{\pi}{8} \int \frac{ \xd^3 \underline{k}}{(2 \pi)^3} \, \frac{t^3}{R^2} |H_{\nu}(k t)|^2 \left( |\eta_1(\underline{k})|^2 + |\eta_2(\underline{k})|^2 - 2 \overline{\eta_2(\underline{k})} \, \eta_1(\underline{k})  \right)  \right).
\ee
The coherent states satisfy the following completeness relation
\be
D^{-1} \int \xd \eta \, \xd \overline{\eta} \, | \psi_{t, \eta} \rangle \langle \psi_{t, \eta}| = I,
\ee
with $I$ being the identity operator and the constant $D$ is given by
\be
D= \int \xd \eta \, \xd \overline{\eta} \, \exp \left( - \frac{\pi}{4} \int \frac{ \xd^3 \underline{k}}{(2 \pi)^3} \, \frac{t^3}{R^2} |H_{\nu}(k t)|^2 |\eta(\underline{k})|^2 \right).
\ee
The characteristic property of coherent states is to remain coherent under the action of the free field propagator,
\be
\psi_{t_2,\eta_2}(\varphi_2) = \int \xD \varphi_1 \, \psi_{t_1,\eta_1}(\varphi_1) \, Z_{[t_1,t_2],0}(\varphi_1, \varphi_2),
\ee
This equation yields the following relation for the complex functions $\eta_1$  and $\eta_2$ defined on the hypersurfaces $t=t_1$ and $t=t_2$ respectively,
\be
\eta_2(\underline{k}) = \frac{t_1^{3/2} H_{\nu}(k t_1)}{t_2^{3/2} H_{\nu}(k t_2)} \, \eta_1(\underline{k}).
\ee 
Hence, the product $\xi(\underline{k}) = t^{3/2} H_{\nu}(k t) \, \eta(\underline{k})$ is preserved under free evolution in the variable $t$. It will be useful to define the interaction representation in terms of the function $\xi$. The coherent state defined as
\be
\psi_{t, \xi}(\varphi) = K_{t, \xi} \, \exp \left( \int \frac{\xd ^3 \underline{x} \, \xd^3 \underline{k}}{(2 \pi)^3} \, \frac{\xi(\underline{k})}{t^{3/2} H_{\nu}(k t)} \, e^{\im \underline{k}\cdot \underline{x}} \varphi(\underline{x}) \right) \psi_{t,0}(\varphi),
\label{eq:cohstint}
\ee
is invariant under free evolution. We will adopt (\ref{eq:cohstint}) as the interaction representation for coherent states. The normalization factor in (\ref{eq:cohstint}) is equal to
\be
K_{t, \xi} = \exp \left( - \frac{\pi}{8 R^2} \int \frac{\xd^3 \underline{k}}{(2 \pi)^3}  \left( \frac{\overline{H_{\nu}(k t)}}{H_{\nu}(k t)} \xi(\underline{k}) \xi(-\underline{k}) + |\xi(\underline{k})|^2 \right) \right).
\label{eq:normcoh}
\ee
Coherent states can be expanded in terms of multiparticle states as
\be
\psi_{t, \xi}(\varphi) =  \exp \left( - \frac{\pi}{8 R^2} \int \frac{ \xd^3 \underline{k}}{(2 \pi)^3} \, |\xi(\underline{k})|^2  \right) \sum_{n=0}^{\infty} \frac{1}{n!} \int \xd ^3 \underline{k}_1 \cdots \int \xd ^3 \underline{k}_n \, \xi(\underline{k}_1) \cdots \xi(\underline{k}_n) \, \psi_{t,\underline{k}_1, \dots, \underline{k}_n}(\varphi),
\label{eq:expmulpar}
\ee
where $\psi_{t,\underline{k}_1, \cdots, \underline{k}_n}$ denotes the $n$-particle state in the interaction picture with quantum numbers $\underline{k}_1, \dots, \underline{k}_n$.
The inner product between a coherent state and a multiparticle state then results in being
\be
\langle \psi_{t,\underline{k}_1, \cdots, \underline{k}_n} |\psi_{t, \xi} \rangle =  
\exp \left( - \frac{\pi}{8 R^2} \int \frac{ \xd^3 \underline{k}}{(2 \pi)^3} \, |\xi(\underline{k})|^2  \right)
\xi(\underline{k}_1) \cdots \xi(\underline{k}_n)  \left(\frac{\pi}{4 R^2}\right)^n.
\label{eq:mulpar}
\ee

\subsection{Asymptotic amplitude}

We compute in this section the amplitude (\ref{eq:ampl}) associated with the spacetime region $[t_1,t_2]$ in the case of the free theory. Considering the state given by the tensor product of two coherent states in the interaction representation defined at times $t_1$ and $t_2$ by the complex functions $\xi_1$ and $\xi_2$ respectively, such that $\psi_{t_1, \xi_1} \otimes \overline{\psi_{t_2, \xi_2}} \in {\cal H}_{t_1} \otimes {\cal H}_{t_2}^*$, where we denote with ${\cal H}_{t_1}$ the Hilbert space associated with the hypersurface $t=t_1$ and with ${\cal H}_{t_2}$ the Hilbert space associated with the hypersurface $t=t_2$ with the same orientation. The amplitude for the non interacting theory is obtained from (\ref{eq:ampl}) and can be interpreted as the transition amplitude from the coherent state $\psi_{t_1, \xi_1}$ to the coherent state $\psi_{t_2, \xi_2}$,
\be
\rho_{[t_1,t_2],0}(\psi_{t_1, \xi_1} \otimes \overline{\psi_{t_2, \xi_2}}) = \int \xD \varphi_1 \, \xD \varphi_2 \, \overline{\psi_{t_2, \xi_2} (\varphi_2)} \, \psi_{t_1, \xi_1}(\varphi_1) \, Z_{[t_1,t_2],0} (\varphi_1, \varphi_2).
\label{eq:freeampl0}
\ee
This amplitude is by construction independent of the initial and final times $t_1$ and $t_2$, and reduces to the inner product between the coherent states $\psi_{t_1, \xi_1}$ and $\psi_{t_2, \xi_2}$,
\be
\rho_{[t_1,t_2],0}(\psi_{t_1, \xi_1} \otimes \overline{\psi_{t_2, \xi_2}})
= \exp \left( - \frac{\pi}{8} \int \frac{\xd^3 \underline{k}}{(2 \pi)^3} \, \frac{1}{R^2}  \left( |\xi_1(\underline{k})|^2 + |\xi_2(\underline{k})|^2 - 2 \overline{\xi_2( \underline{k})} \, \xi_1(\underline{k})  \right)  \right) 
= \langle \psi_{\xi_2} | \psi_{ \xi_1} \rangle.
\label{eq:freeampl}
\ee
We can then trivially take the limit for asymptotic values of the times $t_1$ and $t_2$ and interpret (\ref{eq:freeampl}) as the elements of the S-matrix ${\cal S}_0$ of the free theory,
\be
\langle \psi_{\xi_2}| {\cal S}_0 | \psi_{\xi_1} \rangle = \langle \psi_{\xi_2} | \psi_{ \xi_1} \rangle.
\ee

\section{Region with spacelike boundary - Interacting theory}
\label{sec:int}

We now turn to the study of the interacting theory. We will start in the next section by considering the interaction of the scalar field with a source field confined inside the spacetime region of interest. The amplitude derived in this case will then be used in the subsequent section to express the amplitude for a general interaction by means of functional derivative techniques, following the same approach as in \cite{CoOe:smatrixgbf}

\subsection{Theory with source}

Consider the interaction of the scalar field with a real source field $\mu$ described by the action
\be
S_{[t_1,t_2],\mu}(\phi) = S_{[t_1,t_2],0}(\phi) + \int \xd ^4 x \, \sqrt{-g(x)} \, \mu(x) \phi(x),
\label{eq:actionsrc}
\ee
where $S_{[t_1,t_2],0}$ is the free action (\ref{eq:action}). We assume that the source field $\mu$ vanishes outside the spacetime region $[t_1,t_2]$. 

\subsubsection{Field propagator}

The field propagator corresponding to the action (\ref{eq:actionsrc}) can be evaluate with the same technique applied in Section \ref{sec:qt}, namely shift the integration variable of the path integral by a classical solution of the free theory matching the boundary configurations $\varphi_1$ and $\varphi_2$ on $t=t_1$ and $t=t_2$ respectively. The field propagator then results to be expressed in terms of the free field propagator (\ref{eq:propt}),
\be
Z_{[t_1,t_2],\mu}(\varphi_1, \varphi_2) = \frac{N_{[t_1,t_2],\mu}}{N_{[t_1,t_2],0}} \, Z_{[t_1,t_2],0}(\varphi_1, \varphi_2) \, \exp \left(\im \int \xd ^3 \underline{x} \left( \mu_1(\underline{x}) \varphi_1(\underline{x}) + \mu_2(\underline{x}) \varphi_2(\underline{x}) \right) \right),
\label{eq:propmu}
\ee
where the subscript $\mu$ has been added to the quantity referring to the interacting theory. The quantity $\mu_1$ and $\mu_2$ in the right-hand side of (\ref{eq:propmu}) are defined as
\bea
\mu_1(\underline{x}) &\defeq&  \int_{t_1}^{t_2} \xd t \, \sqrt{-g(t, \underline{x})} \, \frac{\delta_k(t, t_2)}{\delta_k(t_1, t_2)} \, \mu(t, \underline{x}), \\
\mu_2(\underline{x}) &\defeq&  \int_{t_1}^{t_2} \xd t \, \sqrt{-g(t, \underline{x})} \, \frac{\delta_k(t_1, t)}{\delta_k(t_1, t_2)} \, \mu(t, \underline{x}).
\eea
The normalization factor $N_{[t_1,t_2],\mu}$ is formally equal to
\be
N_{[t_1,t_2],\mu} = \int_{\phi|_{t_1}=\phi|_{t_2}=0} \xD \phi \, e^{\im S_{[t_1,t_2],\mu} (\phi)}.
\label{eq:Nmu}
\ee
Again we evaluate the integral by shifting of the integration variable. In this case we shift $\phi$ by the function $\alpha$, solution of the inhomogeneous Klein-Gordon equation
\be
\left[ \frac{t^2}{R^2} \left(\partial_t^2 - \Delta_{\underline{x}} \right) -\frac{2t}{R^2} \partial_t +m^2 \right] \alpha(t, \underline{x}) =\mu(t, \underline{x}),
\ee
with vanishing boundary conditions,
\be
\alpha |_{t=t_1} = \alpha |_{t=t_2}=0.
\ee
The function $\alpha$ results to be
\be
\alpha(x) =  \int_{[t_1,t_2]} \xd^4 x' \, \sqrt{-g(x')} \, G(x,x')\, \mu(x'),
\ee
where $G(x,x')$ is the Green function of the Klein-Gordon equation, with vanishing boundary conditions, given by
\be
G(x,x') =  \frac{\pi}{2 R^2} \int \frac{\xd ^3 \underline{k}}{(2 \pi)^3}  \left( \frac{ \delta_k(t,t_1)\delta_k(t',t_2)}{\delta_k(t_1,t_2)} - \theta(t-t') \, \delta_k(t,t') \right) e^{\im \underline{k}\cdot(\underline{x}-\underline{x}')},
\label{eq:green}
\ee
where $\theta(t)$ is the step function
\be
\theta(t) = \left\{ \begin{matrix} 1 & \mbox{if} \ \ t>0, \\ 0 & \mbox{if} \ \  t<0. \end{matrix} \right.
\ee
Finally the noramlization factor (\ref{eq:Nmu}) can be written as
\be
N_{[t_1,t_2],\mu} = N_{[t_1,t_2],0} \, \exp \left( \frac{\im}{2} \int_{[t_1,t_2]^2} \xd^4 x \, \xd^4 x' \, \sqrt{g(x')g(x)} \, \mu(x) \, G(x,x')\, \mu(x')  \right).
\label{eq:Nmu2}
\ee
Substituting in (\ref{eq:propmu}), we obtain for the field propagator the expression
\begin{align}
Z_{[t_1,t_2],\mu}(\varphi_1, \varphi_2) =&  \, Z_{[t_1,t_2],0}(\varphi_1, \varphi_2) \, \exp \left(\im \int \xd ^3 \underline{x} \left( \mu_1(\underline{x}) \varphi_1(\underline{x}) + \mu_2(\underline{x}) \varphi_2(\underline{x}) \right) \right) \nonumber\\
& \times \exp \left( \frac{\im}{2} \int_{[t_1,t_2]^2} \xd^4 x \, \xd^4 x' \, \sqrt{g(x')g(x)} \, \mu(x) \, G(x,x')\, \mu(x')  \right).
\label{eq:propmu2}
\end{align}
It can be shown that the propagator (\ref{eq:propmu2}) satisfies the composition rule analogue to (\ref{eq:compprop}),
\be
Z_{[t_1, t_3],\mu} (\varphi_1, \varphi_3)  = \int \xD \varphi_2 \, Z_{[t_1, t_2],\mu} (\varphi_1, \varphi_2)  \, Z_{[t_2, t_3],\mu} (\varphi_2, \varphi_3). 
\ee

The unitarity condition (\ref{eq:unitarity}) for the field propagator (\ref{eq:propmu}) reads
\bea
 \int \xD \varphi_2 \, \overline{Z_{[t_1, t_2],\mu} (\varphi_1, \varphi_2)} \, Z_{[t_1, t_2],\mu} (\varphi_1', \varphi_2)
 &=&  \left| \frac{N_{[t_1,t_2],\mu}}{N_{[t_1,t_2],0}}\right|^2 \int \xD \varphi_2 \, \overline{Z_{[t_1, t_2],0} (\varphi_1, \varphi_2)} \, Z_{[t_1, t_2],0} (\varphi_1', \varphi_2) \nonumber\\
&& \times \exp \left(\im \int \xd ^3 \underline{x} \, \mu_1(\underline{x}) (\varphi_1(\underline{x}) - \varphi_1'(\underline{x}))  \right), \nonumber\\
\eea
We notice that the quotient of the normalization factors has modulus one because of the reality of the source field $\mu$ and the Green function (\ref{eq:green}) appearing in the exponential in right hand side of (\ref{eq:Nmu2}). Recalling the result obtained in Section (\ref{sec:freefieldpropagator}), the integral in $\varphi_2$ gives
\be
\int \xD \varphi_2 \, \overline{Z_{[t_1, t_2],\mu} (\varphi_1, \varphi_2)} \, Z_{[t_1, t_2],\mu} (\varphi_1', \varphi_2)
=  \exp \left(\im \int \xd ^3 \underline{x} \, \mu_1(\underline{x}) (\varphi_1(\underline{x}) - \varphi_1'(\underline{x}))  \right) \delta(\varphi_1 - \varphi_1') =   \delta(\varphi_1 - \varphi_1').
\ee
We conclude that the quantum evolution implemented by the field propagator in the presence of a source field is unitary.


\subsubsection{Asymptotic amplitude}
\label{sec:asymampfree1}

The amplitude $\rho_{[t_1,t_2],\mu}$ associated with the transition from the coherent state $\psi_{t_1, \xi_1}$ to the coherent state $\psi_{t_2, \xi_2}$
is
\be
\rho_{[t_1,t_2],\mu}(\psi_{t_1, \xi_1} \otimes \overline{\psi_{t_2, \xi_2}}) = \int \xD \varphi_1 \, \xD \varphi_2 \, \overline{\psi_{t_2, \xi_2} (\varphi_2)} \, \psi_{t_1, \xi_1}(\varphi_1) \, Z_{[t_1,t_2],\mu} (\varphi_1, \varphi_2).
\label{eq:rhosrc0}
\ee
Using the expression (\ref{eq:propmu}) of the field propagator and introducing two new complex functions $\tilde{\xi}_1$ and $\tilde{\xi}_2$ defined as
\bea
\tilde{\xi}_1 (\underline{k}) &\defeq& \xi_1(\underline{k}) + \im \, t_1^{3/2} H_{\nu}(k t_1) \int \xd ^3 \underline{x} \,  e^{-\im \underline{k} \cdot \underline{x}}\,  \mu_1(\underline{x}), \\
\tilde{\xi}_2 (\underline{k}) &\defeq& \xi_2(\underline{k}) - \im \, t_2^{3/2} H_{\nu}(k t_2)  \int \xd ^3 \underline{x}  \,e^{\im \underline{k} \cdot \underline{x}}\,  \mu_2(\underline{x}),
\eea
we can rewrite (\ref{eq:rhosrc0}) in terms of the free amplitude (\ref{eq:freeampl0}) for the coherent states defined by the functions $\tilde{\xi}_1$ and $\tilde{\xi}_2$ in the form
\begin{align}
\rho_{[t_1,t_2],\mu}(\psi_{t_1, \xi_1} \otimes \overline{\psi_{t_2, \xi_2}}) = 
\rho_{[t_1,t_2],0}(\psi_{t_1, \tilde{\xi}_1} \otimes \overline{\psi_{t_2, \tilde{\xi}_2}})
 \,  \frac{ N_{[t_1,t_2],\mu} K_{t_1, \xi_1} \overline{K_{t_2, \xi_2}} }{ N_{[t_1,t_2],0} K_{t_1, \tilde{\xi}_1} \overline{K_{t_2, \tilde{\xi}_2}} }.
\end{align}
With the expressions (\ref{eq:freeampl}), (\ref{eq:Nmu2}) and (\ref{eq:normcoh}), we finally arrive at the following result,
\begin{multline}
\rho_{[t_1,t_2],\mu}(\psi_{t_1, \xi_1} \otimes \overline{\psi_{t_2, \xi_2}}) = 
\rho_{[t_1,t_2],0}(\psi_{t_1, \xi_1} \otimes \overline{\psi_{t_2, \xi_2}})
\exp \left( \frac{\im}{2} \int_{[t_1,t_2]^2} \xd^4 x \, \xd^4 x' \, \sqrt{g(x')g(x)} \, \mu(x) \, G(x,x')\, \mu(x')  \right)\\
 \times
\exp \left(  \im \int_{[t_1,t_2]} \xd ^4 x \, \sqrt{-g(x)} \, \mu(x)\, \hat{\xi}(x) + \frac{\im}{2} \int_{[t_1,t_2]^2} \xd ^4 x \, \xd^4 x' \, \sqrt{g(x')g(x)} \, \mu(x) \, \beta(x,x')\, \mu(x') \right),
\label{eq:rhosrc2}
\end{multline}
where the complex function $\hat{\xi}$ is the complex classical solution of the Klein-Gordon equation determined by $\xi_1$ and $\xi_2$,
\be
\hat{\xi}(x) = \frac{\pi}{4 R^2} \int \frac{\xd ^3 \underline{k}}{(2 \pi)^3} \left(\xi_1(\underline{k}) \, t^{3/2} \,  \overline{H_{\nu}(k t)} \, e^{\im \underline{k} \cdot \underline{x}}+ \overline{\xi_2}(\underline{k}) \, t^{3/2} \,  H_{\nu}(k t) \, e^{-\im \underline{k} \cdot \underline{x}}\right).
\label{eq:hatxi}
\ee
This equation establishes a one-to-one correspondence between pairs of coherent states parametrized by pairs of functions $(\xi_1,\xi_2)$ and complex solutions $\hat{\xi}$ of the Klein-Gordon equation.
The function $\beta$ in the right hand side of (\ref{eq:rhosrc2}) is defined as
\begin{align}
\beta(x,x') = & \, 
\frac{ \pi}{2 R^2 } \int \frac{\xd ^3 \underline{k} }{(2 \pi)^3}  \left(-  \frac{ \delta_k(t,t_1)\delta_k(t',t_2)}{\delta_k(t_1,t_2)} -\frac{1}{2} \delta_k(t',t)+ \frac{\im}{2}(t' t)^{3/2} \left[ J_{\nu}(k t')J_{\nu}(k t)  + Y_{\nu}(k t')Y_{\nu}(k t) \right] \right) \nonumber\\
& \times e^{\im \underline{k} \cdot (\underline{x}- \underline{x}')}.
\end{align}
The combination of the Green function $G$ with the function $\beta$ according to (\ref{eq:rhosrc2}) gives
\bea
G_F(x,x') &\defeq& G(x,x') + \beta(x,x'), \nonumber\\ 
&=& \frac{\im \pi}{4 R^2 } \int \frac{\xd ^3 \underline{k} }{(2 \pi)^3}   \left( 2 \im \theta(t-t') \, \delta_k(t,t') + \im \delta_k(t',t)+ (t' t)^{3/2} \left[ J_{\nu}(k t')J_{\nu}(k t)  + Y_{\nu}(k t')Y_{\nu}(k t) \right] \right)\nonumber\\
&\times& e^{\im \underline{k} \cdot (\underline{x}- \underline{x}')} , \nonumber\\
&=& \frac{\im \pi}{4 R^2 } \int \frac{\xd ^3 \underline{k} }{(2 \pi)^3}    \, (t't)^{3/2} \left( \theta(t-t') \, \overline{H_{\nu}(k t)} \, H_{\nu}(k t') + \theta(t'-t) \, H_{\nu}(k t) \, \overline{H_{\nu}(k t')} \right)e^{\im \underline{k} \cdot (\underline{x}- \underline{x}')}.
\label{eq:Fprop0}
\eea
This expression can be related with the Feynman propagator for the scalar field in Minkowski spacetime via the asymptotic limit defined by (\ref{eq:dStoM}). The limit of (\ref{eq:Fprop0}),
\be
G_F(x,x') =  \im  \int \frac{\xd ^3 \underline{k} }{(2 \pi)^3 2k}\, \left( \theta(t-t') \, e^{\im k (t'-t)} + \theta(t'-t) \, e^{\im k(t -t')} \right)e^{\im \underline{k} \cdot (\underline{x}- \underline{x}')},
\ee
results to be equal to the epxression of the Feynman propagator for a massless scalar field in Minkowski spacetime.
It is possible to evaluate the integral in (\ref{eq:Fprop0}) performing the integration in the angular components of the vector $\underline{k}$
\be
G_F(x,x') = \im \frac{(tt')^{3/2}}{8 \pi R^2} 
\int_{0}^{\infty} \xd k \, k \, \frac{\sin \left( k |\underline{x}-\underline{x}'| \right)}{ |\underline{x}-\underline{x}'|} 
\left(
\theta(t'-t) \, \overline{{H}_{\nu}}(k t') \, {H}_{\nu}(k t) +
\theta(t-t') \, \overline{{H}_{\nu}}(k t) \, {H}_{\nu}(k t') 
\right),
\label{eq:GF1}
\ee
and then using the relation 6.672.3 of \cite{GR:tables} to obtain
\be
G_F(x,x')= - \frac{1}{ 8 \pi R^2}  \frac{1}{\cos(\nu \pi)} \frac{1}{\sqrt{1-p(x,x')^2}} P_{\nu - 1/2}^1 (-p(x,x') + \im 0), \nonumber\\
\ee
where $P$ is the associated Legendre function, and $p(x,x')$ is the de Sitter invariant distance given by
\be
p(x,x')= \frac{t^2+t'^2-|\underline{x}- \underline{x}'|^2}{2 t't}.
\ee
Finally, the function $G_F$ can be expressed in terms of the hypergeometric function $F$,
\be
G_F(x,x') = - \frac{\im}{ 16 \pi R^2} \frac{\left( \frac{1}{4} - \nu^2 \right)}{\cos(\nu \pi)} \,  F \left( -\nu + \frac{3}{2}, \nu +\frac{3}{2}; 2 ; \frac{1+p(x,x')-  \im 0}{2}\right).
\label{eq:Fprop}
\ee
This form of the function $G_F$ coincides with the expression of the Feynman propagator computed in \cite{Schomblond:1976xc} (see formula (A-16) of \cite{Schomblond:1976xc}).
The Feynman propagator (\ref{eq:Fprop}) satisfies the inhomogeneous Klein-Gordon equation,
\be
\left[ \frac{t^2}{R^2} \left(\partial_t^2 - \Delta_{\underline{x}} \right) -\frac{2t}{R^2} \partial_t +m^2 \right] G_F(t,\underline{x},t',\underline{x}') = (-g(t, \underline{x}))^{-1/2} \delta(t-t') \delta^3(\underline{x}-\underline{x}'),
\ee
with the boundary conditions
\bea
G_F(t,\underline{x},t',\underline{x}')\big|_{t=t_1} &=& \frac{\im \pi}{ 4 R^2} \int \frac{\xd ^3 \underline{k} }{(2 \pi)^3}    \, (t't_1)^{3/2}  \, \overline{H_{\nu}(k t')} \, H_{\nu}(k t_1) \, e^{\im \underline{k} \cdot (\underline{x}- \underline{x}')}, \\
G_F(t,\underline{x},t',\underline{x}')\big|_{t=t_2} &=& \frac{\im \pi}{ 4 R^2} \int \frac{\xd ^3 \underline{k} }{(2 \pi)^3}    \, (t't_2)^{3/2}  \, \overline{H_{\nu}(k t_2)} \, H_{\nu}(k t') \, e^{\im \underline{k} \cdot (\underline{x}- \underline{x}')}.
\eea
Finally, the amplitude in the theory with the source field results to be
\begin{multline}
\rho_{[t_1,t_2],\mu}(\psi_{t_1, \xi_1} \otimes \overline{\psi_{t_2, \xi_2}}) = \langle \psi_{ \xi_2} |\psi_{ \xi_1} \rangle  \exp \left( \im \int_{[t_1,t_2]} \xd ^4 x \, \sqrt{-g(x)} \, \mu(x)\, \hat{\xi}(x) \right) \\ 
\times \exp \left(\frac{\im}{2}  \int_{[t_1,t_2]^2} \xd ^4 x \, \xd ^4 x' \, \sqrt{g(x)g(x')} \, \mu(x)\, G_F(x,x') \, \mu(x') \right). 
\label{eq:amplsrc}
\end{multline}
This expression is independent of the times $t_1$ and $t_2$ as long as the source field $\mu$ vanishes outside the region $[t_1,t_2]$. Therefore the limit for asymptotic values of the times $t_1$ and $t_2$ is trivial and we can then remove the restriction on the support of $\mu$. Finally we interpret (\ref{eq:amplsrc}) as the elements of the S-matrix for the theory with the source interaction,
\begin{multline}
\langle \psi_{\xi_2}| {\cal S}_{\mu} | \psi_{\xi_1} \rangle = \langle \psi_{\xi_2} | {\cal S}_{0} |\psi_{ \xi_1} \rangle \exp \left(  \im\int_{[t_1,t_2]} \xd ^4 x \, \sqrt{-g(x)} \, \mu(x)\, \hat{\xi}(x) \right) \\ 
\times \exp \left(\frac{\im}{2}  \int_{[t_1,t_2]^2} \xd ^4 x \, \xd ^4 x' \, \sqrt{g(x)g(x')} \, \mu(x)\, G_F(x,x') \, \mu(x') \right). 
\label{eq:amplsrc1}
\end{multline}

\subsection{General interaction}
\label{sec:genint1}

Consider the action of a scalar field in the presence of an arbitrary potential $V$ vanishing outside the spacetime region $[t_1,t_2]$,
\be
 S_{[t_1,t_2],V}(\phi)=S_{[t_1,t_2],0}(\phi)+\int_{[t_1,t_2]} \xd^4 x\sqrt{-g(x)}\, V(x,\phi(x)).
\ee
Following \cite{CoOe:smatrixgbf} we use the technique of functional derivatives to rewrite the exponential of $\im$ times this action,
\be
\exp \left( \im S_{[t_1,t_2],V}(\phi)\right) = \exp \left(\im \int_{[t_1,t_2]} \xd^4 x\sqrt{-g(x)} \, V\left(x, -\im \frac{\delta}{\delta \mu(x)} \right) \right) \exp \left( \im S_{[t_1,t_2],\mu}(\phi)\right) \bigg|_{\mu=0},
\label{eq:fd}
\ee
where $S_{[t_1,t_2],\mu}$ denotes the action for the theory with the source interaction (\ref{eq:actionsrc}). We can now perform all the calculations of the previous section by moving the functional derivative term in (\ref{eq:fd}) to the front. Then, the amplitude in the presence of the potential $V$ can be immediately written as
\be
\rho_{[t_1,t_2],V}(\psi_{t_1, \xi_1} \otimes \overline{\psi_{t_2, \xi_2}}) = \exp \left(\im \int_{[t_1,t_2]} \xd^4 x\sqrt{-g(x)} \, V\left(x, -\im \frac{\delta}{\delta \mu(x)} \right) \right) \rho_{[t_1,t_2],\mu}(\psi_{t_1, \xi_1} \otimes \overline{\psi_{t_2, \xi_2}})\bigg|_{\mu=0}.
\label{eq:genintasyamp1}
\ee
We notice first that the use of coherent states to compute the asymptotic amplitude is not indispensable and the same expression (\ref{eq:genintasyamp1}) results for generic states $\psi_{t_1,1} \otimes \overline{\psi_{t_2,2}}$. Next, by recalling the independence from times $t_1$ and $t_2$ of the amplitude in the presence of a source interaction, expression (\ref{eq:genintasyamp1}) also does not depend on $t_1$ and $t_2$. This allows us to remove the restriction on the potential $V$. The $S$-matrix elements can then written as
\be
\langle \psi_{2}| {\cal S}_{V} | \psi_{1} \rangle = \exp \left(\im \int \xd^4 x \sqrt{-g(x)} \, V\left(x, -\im \frac{\delta}{\delta \mu(x)} \right) \right) \langle \psi_{2}| {\cal S}_{\mu} | \psi_{1} \rangle \bigg|_{\mu=0}.
\ee

\section{Region with timelike boundary - Free theory}
\label{sec:hypfree}

In this and the next section, we develop the quantum theory for a scalar field in the hypercylinder region defined in Section \ref{sec:hyp}. We will closely follow the treatment of Sections \ref{sec:free} and \ref{sec:int}.

\subsection{Field propagators}
\label{sec:hypfreefieldpropagator}

We start by evaluating the field propagator $Z_{[\varrho_1,\varrho_2],0}$ of the free theory associated with the region $[\varrho_1,\varrho_2]$, namely the region bounded by the hypercylinders of radii $\varrho_1$ and $\varrho_2$. The expression of $Z_{[\varrho_1,\varrho_2],0}$ results from the substitution of the action (\ref{eq:actbound3}) in (\ref{eq:propcl}),
\be
Z_{[\varrho_1, \varrho_2],0}(\varphi_1, \varphi_2) = N_{[\varrho_1, \varrho_2],0} \exp \left( \frac{\im}{2} \int \xd t \, \xd \Omega \, 
\begin{pmatrix}\varphi_1 & \varphi_2 \end{pmatrix} {\cal W}_{[\varrho_1, \varrho_2]}  
\begin{pmatrix} \varphi_1 \\ \varphi_2 \end{pmatrix} \right),
\label{eq:prophyp2}
\ee
where the ${\cal W}_{[\varrho_1, \varrho_2]}$ is a 2x2 matrix with elements given by (\ref{eq:Whyp1}), (\ref{eq:Whyp2}) and (\ref{eq:Whyp3}). The propagator $Z_{[\varrho_1, \varrho_2],0}$ must satisfy a composition rule analogue to (\ref{eq:compprop}),
\be
Z_{[\varrho_1, \varrho_3],0}(\varphi_1, \varphi_3) = \int \xD \varphi_2 \, Z_{[\varrho_1, \varrho_2],0}(\varphi_1, \varphi_2) \, Z_{[\varrho_2, \varrho_3],0}(\varphi_2, \varphi_3).
\ee
As in Section \ref{sec:freefieldpropagator}, this relation fixes the normalization factor appearing in (\ref{eq:prophyp2}),
\be
N_{[\varrho_1, \varrho_2],0} = \det \left( - \frac{\im}{2 \pi}\frac{ R^2}{t^2} \frac{1}{ k \, \Delta_k (\varrho_1, \varrho_2)}\right)^{1/2}.
\label{eq:normfhyp2}
\ee
The quantum evolution implemented by $Z_{[\varrho_1, \varrho_2],0}$ turns out to be unitary since the following relation holds,
\be
\int \xD \varphi_2 \, \overline{Z_{[\varrho_1, \varrho_2],0} (\varphi_1, \varphi_2)} \, Z_{[\varrho_1, \varrho_2],0} (\varphi_1', \varphi_2) = \delta(\varphi_1 - \varphi_1'),
\label{eq:unitarityhyp}
\ee
as can be easily checked using (\ref{eq:normfhyp2}). We now turn to the free field propagator associated with the region enclosed by the hypercylinder of radius $\varrho$. The expression (\ref{eq:propcl}) with the action (\ref{eq:actbound2}) gives
\be
Z_{\varrho,0} (\varphi)= N_{\varrho,0} \exp \left( -\frac{1}{2} \int \xd t \, \xd \Omega \, \varphi(t, \Omega) \, \im \frac{R^2}{t^2} \, \varrho^2 \,k \frac{j_l'(k \varrho)}{j_l(k \varrho)} \, \varphi (t, \Omega)\right).
\label{eq:prophyp1}
\ee
The composition rule satisfied by $Z_{\varrho,0}$ involves the propagator (\ref{eq:prophyp2}),
\be
Z_{\varrho_1,0}(\varphi_1) = \int \xD \varphi_2 \, Z_{\varrho_2,0}(\varphi_2) \, Z_{[\varrho_1, \varrho_2],0}(\varphi_1, \varphi_2).
\ee
This equation translates into the following equality for the normalization factors,
\be
N_{\varrho_1,0} = N_{\varrho_2,0} \, N_{[\varrho_1,\varrho_2],0} \int \xD \varphi_2 \, \exp \left( \frac{\im}{2} \int \xd t \, \xd \Omega \, \frac{R^2}{t^2}  \, \varphi_2(t, \Omega)  \left( \frac{1}{k \, \Delta_k(\varrho_1,\varrho_2)} \frac{j_l(k \varrho_1)}{j_l(k \varrho_2)} \, \varphi_2 \right) (t, \Omega) \right).
\ee
Using (\ref{eq:normfhyp2}), we find the solution,
\be
N_{\varrho_1,0} =\det \left( \frac{1}{ j_l(k \varrho_1)}\right)^{1/2}.
\label{eq:normfhyp1}
\ee
We conclude this subsection by looking at the asymptotic limit of the field propagators (\ref{eq:prophyp2}) and (\ref{eq:prophyp1}). We expect to recover the corresponding field propagators defined in Minkowski spacetime where the notion of the hypercylinder was originally introduced, in particular we refer to Section IV.A.2 of \cite{CoOe:smatrixgbf}. The comparison between de Sitter propagators (\ref{eq:prophyp2}) and (\ref{eq:prophyp1}), and Minkowski propagators (74) and (75) of \cite{CoOe:smatrixgbf}, manifests only one significant difference: The presence of the quotient $\frac{R^2}{t^2}$ in the expression of the former propagators. We notice that it is precisely this quotient that get canceled in the limit (\ref{eq:dStoM}), provided that $k$ in (\ref{eq:prophyp2}) and (\ref{eq:prophyp1}) equals $\sqrt{E^2-m^2}$ (for $E^2>m^2$) appearing in the propagators (74) and (75) of \cite{CoOe:smatrixgbf}. This identification is consequently valid only for positive $k$. An alternative would be to replace $k$ with $|k|$. Such change will receives its justification in the next section, where we introduce the vacuum state.


\subsection{Vacuum state}
\label{sec:vacuumhyp}

The Gaussian ansatz for the vacuum wave function associated with the hypercylinder of radius $\varrho$ reads
\be
\psi_{\varrho,0}(\varphi)=C_{\varrho}  \exp\left(-\frac{1}{2}\int \xd t \, \xd \Omega \,  \varphi(t, \Omega)(A_{\varrho} \varphi)(t, \Omega)\right) ,
\ee
where $C_{\varrho}$ is the normalization factor and $A_{\varrho}$ denotes a family of operators indexed by ${\varrho}$. Applying the same techniques developed in \cite{Co:vacuum}, we obtain the following form for the vacuum wave function,
\be
 \psi_{\varrho,0}(\varphi) = C_{\varrho}  \exp\left(\frac{\im}{2}\int \xd t \, \xd \Omega \, \frac{R^2}{t^2} \varrho^2 \,  \varphi(t, \Omega) |k| \frac{h_{l}'(|k| \varrho)}{h_{l}(|k| \varrho)} \varphi(t, \Omega)\right),
\label{eq:vachyp}
\ee 
where $h_l$ is the spherical Bessel function of the third kind, $h_l = j_l + \im n_l$, and the appearance of the modulus of $k$ guarantees the normalizability of the vacuum state, as can be seen from the analytic continuation of the spherical Bessel functions (see in particular formulas 10.1.34-35 of \cite{AbSt:handbook}). Based on these properties and the observation at the end of the previous section concerning the asymptotic limit of the field propagator, in the rest of paper the spherical Bessel functions will depend on $|k|$ and not $k$.
$C_{\varrho}$ is the normalization factor given by (up to a phase)
\bea
|C_{\varrho}|^{-2} = \int \xD \varphi \, \exp \left( - \frac{1}{2} \int \xd t \, \xd \Omega \, \varphi(t, \Omega) \, \frac{R^2}{t^2} \frac{1}{|k| \, |h_l(|k| \varrho)|^2} \, \varphi(t, \Omega)\right) = \det \left( \frac{R^2}{2 \pi \, t^2 \, |k| \, |h_l(|k| \varrho)|^2}\right)^{-1/2}.
\eea
As in Section \ref{sec:vacuum}, the phase of $C_{\varrho}$ is fixed by the invariance of the vacuum state under free evolution,
\be
\psi_{\varrho_2,0}(\varphi_2) = \int \xD \varphi_1 \, \psi_{\varrho_1,0}(\varphi_1) \, Z_{[\varrho_1,\varrho_2],0}(\varphi_1, \varphi_2),
\ee
which implies the following identity for the normalization factors
\bea
C_{\varrho_2} &=& C_{\varrho_1} \, N_{[\varrho_1,\varrho_2],0} \, \int \xD \varphi_1 \exp \left( - \frac{1}{2} \int \xd t \, \xd \Omega \, \frac{R^2}{t^2} \, \varphi_1 \left[- \frac{\im}{|k| \Delta_k(\varrho_1,\varrho_2)} \,  \frac{h_{l}(|k| \varrho_2)}{  h_{l}(|k| \varrho_1)} \right] \varphi_1 \right), \nonumber\\
&=& C_{\varrho_1} \,  \det \left(  \frac{h_{l}(|k| \varrho_2)}{  h_{l}(|k| \varrho_1)}  \right)^{-1/2},
\eea
where (\ref{eq:normfhyp2}) has been used. It can be easily verify that the above equation admits the solution
\be
C_{\varrho} = \det \left( \frac{R}{t \sqrt{2 \pi \, |k|} \, h_l(|k| \varrho)}\right)^{1/2}.
\ee
The asymptotic limit of (\ref{eq:vachyp}) coincides with the vacuum wave function defined on the hypercylinder in Minkowski spacetime if $|k|=\sqrt{E^2-m^2}$ (for $E^2>m^2$), see (79) and (80) of \cite{CoOe:smatrixgbf}. 

Now that we have at our disposal both their asymptotic limit, we can compare the vacuum state defined on the hypersurface of constant time $t$, (\ref{eq:vac}), with the one on the hypercylinder of radius $\varrho$, (\ref{eq:vachyp}). This amounts to compare the corresponding asymptotic Minkowski vacuum wave functions, which are related at spatial and temporal infinity as explained in \cite{CoOe:smatrixgbf}. We then conclude that the same relation holds for the vacua (\ref{eq:vac}) and (\ref{eq:vachyp}).

The next step is the definition of coherent states on the hypercylinder.


\subsection{Coherent states}

The wave function of a coherent state defined on the hypercylinder of radius $\varrho$, in Schr\"odinger representation, is parametrized by a complex function $\eta$,
\be
\psi_{\varrho, \eta}(\varphi) = K_{\varrho, \eta} \, \exp \left( \int \xd t \, \xd \Omega \, \xd k \sum_{l,m} \eta_{l,m}(k) \, 
|t|^{-1/2} \overline{{\mathscr H}_{\nu}(kt) } Y_{l}^{-m}(\Omega) \, \frac{|k|}{4}  
\varphi(t, \Omega) \right) \psi_{\varrho,0}(\varphi),
\ee
where the integration in $k$ and the sum over the indexes $l$ and $m$ are defined as in (\ref{eq:classsolhyp}). The normalization factor $K_{\varrho, \eta}$ results to be
\bea
K_{\varrho, \eta} 
&=& \exp \left(- \frac{1}{16 \, R^2} \int
 \xd k \sum_{l,m} \, k^2 |h_l(|k| \varrho)|^2 \left[ |\eta_{l,m}(k)|^2 - \eta_{l,m}(k) \eta_{l,-m}(-k) \right]  \right).
\eea
The inner product of two coherent states defined by the complex functions $\eta$ and $\eta'$ results to be
\be
\langle \psi_{\varrho, \eta'} | \psi_{\varrho, \eta} \rangle = \exp \left(- \frac{1}{16 \, R^2} \int
 \xd k \sum_{l,m} \, k^2 |h_l(|k| \varrho)|^2 \left[ |\eta_{l,m}(k)|^2 + |\eta_{l,m}'(k)|^2 - 2 \eta_{l,m}(k) \overline{\eta_{l,m}'(k)} \right]  \right).
\ee
The completeness relation satisfied by the coherent states satisfy can be written as
\be
D^{-1} \int \xd \eta \, \xd \overline{\eta} \, | \psi_{\varrho, \eta} \rangle \langle \psi_{\varrho, \eta}| = I,
\label{eq:complrelhyp}
\ee
with $I$ being the identity operator and the constant $D$ is given by
\be
D= \int \xd \eta \, \xd \overline{\eta} \, \exp \left(  - \frac{1}{8 R^2} \int \xd k \sum_{l,m}  \, k^2 \, |h_{l}(|k| \varrho)|^2 \, |\eta_{l,m}(k)|^2 \right).
\ee

Two coherent states parametrized by the complex functions $\eta_1$ and $\eta_2$, defined respectively on the hypercylinders of radii $\varrho_1$ and $\varrho_2$ are related by
\be
\psi_{\varrho_2,\eta_2}(\varphi_2) = \int \xD \varphi_1 \, \psi_{\varrho_1,\eta_1}(\varphi_1) \, Z_{[\varrho_1,\varrho_2],0}(\varphi_1, \varphi_2).
\ee
This equality holds if the complex functions $\eta_1$ and $\eta_2$ satisfy
\be
\eta_{1,l,m}(k) = \eta_{2,l,m}(k) \frac{h_{l}(|k| \varrho_2)}{h_{l}(|k| \varrho_1)}.
\ee
As a consequence, the function $\xi_{l,m}(k) = \eta_{l,m}(k) h_{l}(|k| \varrho)$ appears to be independent of the radius $\varrho$, since it is preserved under free evolution. The interaction picture for coherent states will then be defined in terms of $\xi$, namely
\be
\psi_{\varrho, \xi}(\varphi) = K_{\varrho, \xi} \, \exp \left( \int \xd t \, \xd \Omega \, \xd k \sum_{l,m} \xi_{l,m}(k) \frac{ t^{-1/2} \overline{{\mathscr H}_{\nu}(kt) } Y_{l}^{-m}(\Omega)}{h_l(|k| \varrho)} \,  \varphi(t, \Omega) \right) \psi_{\varrho,0}(\varphi).
\ee
The normalization factor $K_{\varrho, \xi}$ is equal to
\be
K_{\varrho, \xi} = \exp \left( - \frac{1}{16 R^2} \int \xd k \sum_{l,m} \, k^2  \left( |\xi_{l,m}(k)|^2 - \frac{\overline{h_{l}(|k| \varrho)}}{h_{l}(|k| \varrho)} \xi_{l,m}(k) \xi_{l,-m}(-k) \right)  \right).
\label{eq:normcohhyp}
\ee
The completeness relation (\ref{eq:complrelhyp}) in the interaction picture reads
\be
\tilde{D}^{-1} \int \xd \xi \, \xd \overline{\xi} \, | \psi_{\varrho, \xi} \rangle \langle \psi_{\varrho, \xi}| = I,
\label{eq:complrelhyp2}
\ee
where the constant $\tilde{D}$ is 
\be
\tilde{D}= \int \xd \xi \, \xd \overline{\xi} \, \exp \left(  - \frac{1}{8 R^2} \int \xd k \sum_{l,m} \, k^2 \, |\xi_{l,m}(k)|^2 \right).
\label{eq:Dtilde}
\ee

The expansion of a coherent state in terms of multiparticle states reads
\bea
\psi_{\varrho, \xi}(\varphi) &=& \exp \left(  - \frac{1}{16 R^2} \int \xd k \sum_{l,m} k^2 \, |\xi_{l,m}(k)|^2 \right) \sum_{n=0}^{\infty} \frac{1}{n!} \nonumber\\
& \times & \int \xd k_1 \sum_{l_1,m_1} \cdots \int \xd k_n \sum_{l_n,m_n} \xi_{l_1,m_1}(k_1) \cdots \xi_{l_n,m_n}(k_n) \, \psi_{\varrho, (k_1,l_1,m_1), \dots, (k_n,l_n,m_n)} (\varphi).
\label{eq:expmulparhyp}
\eea
The state with $n$ particles with quantum numbers $(k_1,l_1,m_1), \dots, (k_n,l_n,m_n)$ has been denoted by $\psi_{\varrho, (k_1,l_1,m_1), \dots, (k_n,l_n,m_n)}$ in the interaction picture. The inner product between a coherent state and an $n$-particle state is
\bea
\langle \psi_{\varrho, (k_1,l_1,m_1), \dots, (k_n,l_n,m_n)} |\psi_{\varrho, \xi} \rangle &=& 
\exp \left(  - \frac{1}{16 R^2} \int \xd k \sum_{l,m} k^2 \, |\xi_{l,m}(k)|^2 \right) \nonumber\\
& \times &
\xi_{l_1,m_1}(k_1) \cdots \xi_{l_n,m_n}(k_n)  \frac{k_1^{2}}{4 R^{2}} \cdots \frac{k_n^{2}}{4 R^{2}}.
\label{eq:mulparhyp}
\eea


\subsection{Asymptotic amplitude}
\label{sec:asyamplfree}

We now have all the ingredients to evaluate the amplitude associated with the spacetime region enclosed by the hypercylinder of radius $\varrho$, for the coherent state $\psi_{\varrho, \xi}$ defined on the hypercylinder, in the case of the free theory. According to (\ref{eq:ampl}), this amplitude is
\bea
\rho_{\varrho,0}(\psi_{\varrho, \xi} ) &=& \int \xD \varphi \, \psi_{\varrho, \xi}(\varphi) \, Z_{\varrho,0} (\varphi), \nonumber\\
&=& \exp \left( - \frac{1}{16 R^2} \int \xd k \sum_{l,m}  \, k^2  \left[ |\xi_{l,m}(k)|^2  + \xi_{l,m}(k) \, \xi_{l,-m}(-k)  \right]  \right).
\label{eq:freeamplhyp}
\eea
This amplitude is manifestly independent of the radius $\varrho$ of the hypercylinder.

\section{Region with timelike boundary - Interacting theory}
\label{sec:hypint}

The study of the interacting Klein-Gordon theory in the spacetime region bounded by the hypercylinder will parallel the analysis performed in Section \ref{sec:int}. We turn first to the theory describing the interaction with an external source field and we then conclude the section by the treatment of the general interacting theory.

\subsection{Theory with source}

The action of a scalar field interacting with a real source field $\mu$ has the form
\be
S_{\varrho,\mu} (\phi)= S_{\varrho,0} (\phi) + \int \xd ^4 x \sqrt{-g(x)} \, \mu(x) \, \phi(x),
\ee
where $S_{\varrho,0}$ is the action of the free theory (\ref{eq:actbound2}). The source field $\mu$ is assumed to vanish outside the spacetime region $\varrho$, namely $\mu(t,r,\Omega)\big|_{r \geq \varrho} =0$.

\subsubsection{Field propagator}

The field propagator can be expressed in terms of the field propagator of the free theory,
\be
Z_{\varrho,\mu}(\varphi) = \frac{N_{\varrho,\mu}}{N_{\varrho,0}} \, Z_{\varrho,0}(\varphi) \, \exp \left( \im \int \xd t \, \xd \Omega  \, M_l(t,\Omega) \, \frac{1}{j_l(|k| \varrho)} \, \varphi(t, \Omega)\right),
\label{eq:propsrchyp2}
\ee
where we used the expression of the field $\phi$ in terms of the boundary field configuration $\varphi$, given by (\ref{eq:hypbound}), and we introduced the quantity
\be
M_l(t, \Omega) =  \int_0^{\infty} \xd r \, \sqrt{-g(t,r,\Omega)} \, j_l(|k| r) \, \mu(t,r,\Omega).
\ee
The quotient of normalization factor appearing in the right hand side of (\ref{eq:propsrchyp2}) can be rewritten as follows,
\be
\frac{N_{\varrho,\mu}}{N_{\varrho,0}} = \exp \left( \frac{\im}{2} \int \xd ^4 x \sqrt{-g(x)} \, \alpha(x) \, \mu(x)\right),
\ee
where the function $\alpha$ is a solution of the inhomogeneous Klein-Gordon equation
\be
\left( \frac{t^2}{R^2} \left[ \partial_t^2 - \Delta_r - \Delta_{\Omega}\right] -\frac{2 t}{R^2} \partial_t + m^2 \right)\alpha(t,r, \Omega) = \mu(t,r, \Omega),
\label{eq:inKGhyp}
\ee
with vanishing boundary condition,
\be
\alpha \big|_{r= \varrho} = 0.
\label{eq:bchyp}
\ee
We express the solution of (\ref{eq:inKGhyp}) with the boundary condition (\ref{eq:bchyp}) in terms of the Green function $G$ of the Klein-Gordon equation,
\be
G(x,x') = \frac{1}{2 R^2} \int \xd k \sum_{l,m} k^2 \, (tt')^{3/2} {\mathscr H}_{\nu}(k t) \overline{{\mathscr H}_{\nu}(kt') } Y_l^{m}(\Omega) \overline{Y_l^{m}(\Omega')} \, g_{l}(k,r,r'),
\ee
where the function $g_l$ is given by
\be
g_l(k,r,r') = \theta(r-r')[ j_l(|k| r) \, n_l(|k| r')- n_l(|k| r) \, j_l(|k| r')] - j_l(|k| r) \, n_l(|k| r')+ j_l(|k| r) \, \frac{n_l(|k| \varrho)}{j_l(|k| \varrho)}\, j_l(|k| r').
\ee
The function $\alpha$ can then be written as
\be
\alpha(x) = \int \xd ^4 x' \sqrt{-g(x')} \, G(x,x') \, \mu(x').
\ee
By substituting this result in (\ref{eq:propsrchyp2}), the field propagator takes the form
\be
Z_{\varrho,\mu}(\varphi) = Z_{\varrho,0}(\varphi)  \exp \left( \im \int \xd t \, \xd \Omega  \, M_l(t,\Omega) \frac{1}{j_l(|k| \varrho)}  \varphi(t, \Omega)\right) \exp \left( \frac{\im}{2} \int \xd ^4 x \, \xd ^4 x'\sqrt{g(x) g(x')} \, \mu(x)  G(x,x')  \mu(x')\right).
\label{eq:propsrchyp}
\ee

\subsection{Asymptotic amplitude}

The amplitude $\rho_{\varrho,\mu}$ associated with the spacetime region $\varrho$, for the coherent state $\psi_{\varrho,\xi}$ defined on the boundary of this region is
\be
\rho_{\varrho,\mu}(\psi_{\varrho,\xi}) = \int \xD \varphi \, \psi_{\varrho,\xi}(\varphi) \, Z_{\varrho,\mu}(\varphi).
\label{eq:amplsrchyp1}
\ee
Introducing the function $\tilde{\xi}$, given by
\be
\tilde{\xi}(t, \Omega) \defeq \int \xd k \sum_{l} \left( \sum_m \xi_{l,m}(k) \frac{ t^{-1/2} \overline{{\mathscr H}_{\nu}(kt)} Y_{l}^{-m}(\Omega)}{h_l(|k| \varrho)} + \frac{\im}{j_l(|k| \varrho)} \, M_l(t,\Omega) \right),
\ee
we can define the new coherent state $\psi_{\varrho, \tilde{\xi}}$. The amplitude (\ref{eq:amplsrchyp1}) can then be expressed in terms of the free amplitude (\ref{eq:freeamplhyp}) for the new coherent state defined by the function $\tilde{\xi}$, 
\be
\rho_{\varrho,\mu}(\psi_{\varrho,\xi}) = \rho_{\varrho,0}(\psi_{\varrho,\tilde{\xi}}) \, \frac{K_{\varrho,\xi}}{K_{\varrho,\tilde{\xi}}} \, 
\exp \left( \frac{\im}{2} \int \xd ^4 x \, \xd ^4 x' \sqrt{g(x) g(x')}\, \mu(x) \, G(x,x') \, \mu(x')\right).
\ee
Substituting the expressions (\ref{eq:freeamplhyp}) and (\ref{eq:normcohhyp}) we arrive at
\begin{multline}
\rho_{\varrho,\mu}(\psi_{\varrho,\xi}) = \rho_{\varrho,0}(\psi_{\varrho,\xi}) \, \exp \left( \im \int \xd ^4 x \sqrt{-g(x)} \, \mu(x) \, \hat{\xi}(x)\right) \\ 
\times  \exp \left( \frac{\im}{2} \int \xd ^4 x \, \xd ^4 x' \sqrt{g(x) g(x')} \, \mu(x)  [G(x,x') + \beta(x,x')] \mu(x') 
\right),
\label{eq:amplsrchyp3}
\end{multline}
where the function $\hat{\xi}$ is the complex classical solution of the Klein-Gordon equation parametrized by $\xi$,
\be
\hat{\xi}(x) = - \frac{1}{8 R^2} \int \xd k \sum_{l,m} j_l(|k| r) \, k^2 \left( \xi_{l,m}(k) \, \overline{Y_l^m(\Omega)} \, |t|^{3/2} \overline{{\mathscr H}_{\nu}}(k t)  -  \xi_{l,-m}(-k) \, Y_l^m(\Omega) \, |t|^{3/2} {\mathscr H}_{\nu}(k t) \right),
\label{eq:hatxihyp}
\ee
implying a one-to-one correspondence between coherent states parametrized by functions $\xi$ and complex solutions $\hat{\xi}$.
The function $\beta$ in the right hand side of (\ref{eq:amplsrchyp3}) is equal to
\be
\beta(x,x') = \frac{\im}{4 R^2} \int \xd k \sum_{l,m} k^2 \, (tt')^{3/2} {\mathscr H}_{\nu}(k t) \overline{{\mathscr H}_{\nu}(kt') } Y_l^m(\Omega) \overline{Y_l^m(\Omega')} \, j_l(|k| r) \frac{h_l(|k| \varrho)}{j_l(|k| \varrho)} j_l(|k| r').
\ee
We denote by $G_F$ the sum of the Green $G$ and the function $\beta$ that appears in the last exponential of (\ref{eq:amplsrchyp3}),
\begin{flalign}
G_F(x,x') \defeq& \, \,  G(x,x') + \beta(x,x'),& & \nonumber\\
=& \, \, 
\frac{\im}{4 R^2} \int \xd k \sum_{l,m} k^2 \,  (tt')^{3/2}  {\mathscr H}_{\nu}(k t) \overline{{\mathscr H}_{\nu}(kt') } \, Y_l^m(\Omega) \overline{Y_l^m(\Omega')} & & \nonumber
\end{flalign}
\vspace{-8mm}
\begin{flalign}
& & \times \left( \theta(r-r') \, j_l(|k| r') \, h_l(|k| r) + \theta(r'-r) \, j_l(|k| r) \, h_l(|k| r') \right) &.
\label{eq:GreenFhyp}
\end{flalign}
This function solves the inhomogeneous Klein-Gordon equation, 
\be
\left( \frac{t^2}{R^2} \left[ \partial_t^2 - \Delta_r - \Delta_{\Omega}\right] -\frac{2 t}{R^2} \partial_t + m^2 \right) G(t,r,\Omega,t',r',\Omega') = (-g(t,r,\Omega))^{-1/2} \delta(t-t')\delta(r-r')\delta(\Omega-\Omega'),
\ee
with the boundary condition
\be
G(t,r,\Omega,t',r',\Omega')\bigg|_{r = \varrho} = \frac{\im}{4 R^2} \int_0^{\infty} \xd k \sum_{l,m} k^2 \,  (tt')^{3/2}  {\mathscr H}_{\nu}(k t) \overline{{\mathscr H}_{\nu}(kt') } \, Y_l^m(\Omega) \overline{Y_l^m(\Omega')} j_l(|k| r') h_l(|k| \varrho).
\label{eq:bcGreenFhyp}
\ee
The equivalence of (\ref{eq:GreenFhyp}) with the Feynman propagator $G_F$ obtained in Section \ref{sec:asymampfree1} is shown in Appendix \ref{sec:apB}.
Finally, we write the amplitude $\rho_{\varrho,\mu}$ as
\begin{multline}
\rho_{\varrho,\mu}(\psi_{\varrho,\xi}) = \rho_{\varrho,0}(\psi_{\varrho,\xi}) \, \exp \left( \im \int \xd ^4 x \sqrt{-g(x)} \, \mu(x) \, \hat{\xi}(x)\right) \\ 
\times  \exp \left( \frac{\im}{2} \int \xd ^4 x \, \xd ^4 x' \sqrt{g(x) g(x')} \, \mu(x)  \,G_F(x,x')\, \mu(x') 
\right).
\label{eq:amplsrchyp4}
\end{multline}
Based on the result of Section \ref{sec:asyamplfree}, no dependence on the radius $\varrho$ of the hypercylinder appears in the amplitude $\rho_{\varrho,\mu}$. Consequently, expression (\ref{eq:amplsrchyp4}) coincides with the asymptotic amplitude for $\varrho \rightarrow \infty$.



\subsection{General interaction}
\label{sec:genint2}

The asymptotic amplitude in the presence of a general interaction can be computed by means of the same functional derivatives techniques used in Section \ref{sec:genint1}. In the present case, we require that the potential $V$ defining the interaction vanishes outside the hypercylinder region, 
\be
V((t,r,\Omega), \phi(t,r, \Omega)) = 0, \qquad \text{if $r \geq \varrho$}.
\ee
The computation of the amplitude for a generic $\psi_{\varrho}$ state on the hypercylinder gives the following result
\be
\rho_{\varrho, V}(\psi_{\varrho}) = \exp \left(\im \int \xd^4 x \sqrt{-g(x)} \, V\left(x, -\im \frac{\delta}{\delta \mu(x)} \right) \right)
\rho_{\varrho, \mu}(\psi_{\varrho}) \bigg|_{\mu=0}.
\label{eq:genintasyamp2}
\ee
The amplitude in the presence of a source interaction, $\rho_{\varrho, \mu}$, does not depend on the radius of the hypercylinder. We can then lift the restriction on $V$, and interpret (\ref{eq:genintasyamp2}) as the asymptotic amplitude for the general interacting theory.

\section{Isomorphism}
\label{sec:iso}

Having in mind previous results obtained applying the GBF to the theory of quantum fields in Minkowski spacetime \cite{CoOe:spsmatrix,CoOe:smatrixgbf} and Euclidean spacetime \cite{CoOe:smatrix2d}, the similarity between the asymptotic amplitudes in the presence of a source field in the two geometries considered, namely, expressions (\ref{eq:amplsrc1}) and (\ref{eq:amplsrchyp4}), should not surprise us. In particular, based on the cited works and on the appearance of the same Feynman propagator in these amplitudes, we expect the two amplitudes to be equivalent under the action of an isomorphism between the corresponding spaces of states. To be more precise the isomorphism ${\cal H}_{t_1} \otimes {\cal H}_{t_2}^* \rightarrow {\cal H}_{\varrho}$, where ${\cal H}_{\varrho}$ denotes the Hilbert space associated with the hypercylinder, is determined by the identification of the complex solutions of the Klein-Gordon equation (\ref{eq:hatxi}) and (\ref{eq:hatxihyp}). Indeed these solutions establish a one-to-one correspondence between complex classical solutions in spacetime and coherent states in ${\cal H}_{t_1} \otimes {\cal H}_{t_2}^*$ and ${\cal H}_{\varrho}$ respectively. Hence, for a given complex classical solution we implement the correspondence $\psi_{t_1,\xi_1} \otimes \overline{\psi_{t_2,\xi_2}} \mapsto \psi_{\varrho,\xi}$ that fixes the isomorphism. The equivalence of the amplitudes (\ref{eq:amplsrc1}) and (\ref{eq:amplsrchyp4}) is then realized if the free amplitudes (\ref{eq:freeampl}) and (\ref{eq:freeamplhyp}) coincide under the action of the isomorphism.

Identifying the complex classical solutions (\ref{eq:hatxi}) and (\ref{eq:hatxihyp}) leads to the following relations for the modes $\xi_{l,m}(k)$ and $\xi_{1,2}(\underline{k})$,
\be
\xi_{l,m}(k) =  e^{\im \nu \pi /2}\frac{ \im^l}{2 \pi} \int \xd \Omega_k \, \xi_1(\underline{k}) \, Y_l^m(\Omega_k),\qquad
\xi_{l,-m}(-k) = - e^{-\im \nu \pi /2} \frac{(- \im )^l}{2 \pi} \int \xd \Omega_k \, \overline{\xi_2(\underline{k})} \, \overline{Y_l^m}(\Omega_k).
\label{eq:isom}
\ee
where the angular coordinates $\Omega_k$ determine the direction of the vector $\underline{k}$. We substitute (\ref{eq:isom}) in the expression of the free amplitude associated with the hypercylinder (\ref{eq:freeamplhyp}),
\bea
\rho_{\varrho,0}(\psi_{\varrho, \xi} ) &=& 
 \exp \left( - \frac{1}{64 \pi^2 R^2} \int_{0}^{\infty} \xd k \int \xd \Omega_k \int \xd \Omega_k' \sum_{l,m}  k^2 \,  \overline{Y_l^m(\Omega_k)} \, Y_l^m(\Omega_k') \right. \nonumber\\
&& \times \left( \overline{\xi_{1}(\underline{k})} \,\xi_{1}(\underline{k}') + \overline{\xi_{2}(\underline{k})} \, \xi_{2}(\underline{k}')  - 2 \overline{\xi_{2}(\underline{k})} \, \xi_{1}(\underline{k}')    \right) \bigg).
\eea
The sum over $l,m$ is performed using the completeness relation of spherical harmonics, see formula (B.88) of \cite{Mes:qm1},
\be
\rho_{\varrho,0}(\psi_{\varrho, \xi} ) =\exp \left( - \frac{\pi}{8  R^2} \int \frac{\xd^3 \underline{k}}{(2 \pi)^3}  \left( |\xi_{1}(\underline{k})|^2 + |\xi_{2}(\underline{k})|^2  - 2 \overline{\xi_{2}(\underline{k})} \, \xi_{1}(\underline{k})    \right) \right) = \rho_{[t_1,t_2],0}(\psi_{t_1,\xi_1} \otimes \overline{\psi_{t_2,\xi_2}} ).
\ee
This result proves the equivalence of the free amplitudes (\ref{eq:freeampl}) and (\ref{eq:freeamplhyp}).

The map between coherent states $\psi_{t_1,\xi_1} \otimes \overline{\psi_{t_2,\xi_2}} \mapsto \psi_{\varrho,\xi}$ implemented by the isomorphism ${\cal H}_{t_1} \otimes {\cal H}_{t_2}^* \rightarrow {\cal H}_{\varrho}$ leads to a map between multiparticle states. We are interested here in the expression of a multiparticle state defined in one setting in terms of multiparticle states defined in the other setting. Consider a state with $n$ particles in ${\cal H}_{t_1} \otimes {\cal H}_{t_2}^*$ with $q$ incoming particles with quantum numbers $\underline{k}_1, \dots, \underline{k}_q$ and $n-q$ outgoing particles with quantum numbers $\underline{k}_{q+1}, \dots, \underline{k}_n$, denoted as
\be
\psi_{\underline{k}_1, \cdots, \underline{k}_q|\underline{k}_{q+1}, \cdots, \underline{k}_n} = | \psi_{\underline{k}_1, \cdots, \underline{k}_q} \rangle  \otimes \langle \psi_{\underline{k}_{q+1}, \cdots, \underline{k}_n} |,
\label{eq:npart}
\ee
where $| \psi_{\underline{k}_1, \cdots, \underline{k}_q} \rangle$ is $q$-particle state introduced in (\ref{eq:mulpar}). We want to evaluate the inner product between the state (\ref{eq:npart}) and the $n$-particle state in ${\cal H}_{\varrho}$ with quantum numbers $(k_1,l_1,m_1), \dots, (k_n,l_n,m_n)$ introduced in (\ref{eq:mulparhyp}). Inserting the completeness relation of coherent states (\ref{eq:complrelhyp2}), 
\be
\langle \psi_{\underline{k}_1, \cdots, \underline{k}_q|\underline{k}_{q+1}, \cdots, \underline{k}_n}| \psi_{(k_1,l_1,m_1), \dots, (k_n,l_n,m_n)} \rangle = \tilde{D}^{-1} \int \xd \xi \, \xd \overline{\xi} \langle \psi_{\underline{k}_1, \cdots, \underline{k}_q|\underline{k}_{q+1}, \cdots, \underline{k}_n}| \psi_{\xi} \rangle \langle \psi_{\xi} | \psi_{(k_1,l_1,m_1), \dots, (k_n,l_n,m_n)} \rangle.
\ee
and using (\ref{eq:mulpar}), (\ref{eq:mulparhyp}), (\ref{eq:Dtilde}) and the relations (\ref{eq:isom}) we perform the integration in $\xd \xi \, \xd \overline{\xi}$ and obtain,
\begin{multline}
\langle \psi_{\underline{k}_1', \cdots, \underline{k}_q'|\underline{k}_{q+1}', \cdots, \underline{k}_n'}| \psi_{(k_1,l_1,m_1), \dots, (k_n,l_n,m_n)} \rangle = (-1)^{n-q} \left( \frac{2 \pi^2}{R^2} \right)^n \, \im^{l_{q+1} + \cdots + l_n- l_1- \cdots -l_q} \\
\times Y_{l_1}^{-m_1}(\Omega_{k_1}) \cdots Y_{l_n}^{-m_n}(\Omega_{k_n})
 \, \delta(k_1'-k_1) \cdots \delta(k_q'-k_q) \delta(k_{q+1}'+k_{q+1}) \cdots \delta(k_n'+k_n).
\end{multline}
We can now write an $n$-particle state in ${\cal H}_{\varrho}$ as a linear combination of $n$-particle states in ${\cal H}_{t_1} \otimes {\cal H}_{t_2}^*$,
\begin{align}
\psi_{\underline{k}_1, \cdots, \underline{k}_q|\underline{k}_{q+1}, \cdots, \underline{k}_n} =& \, 
(-1)^{n-q} \, \frac{(16 \pi^3)^n}{k_1^2 \cdots k_n^2} \, \sum_{l_1,m_1} \cdots \sum_{l_n,m_n} \, \im^{-l_{q+1} - \cdots - l_n+ l_1+ \cdots +l_q} \, 
Y_{l_1}^{m_1}(\Omega_{k_1}) \cdots Y_{l_n}^{m_n}(\Omega_{k_n}) \nonumber\\ 
& \times \psi_{(k_1,l_1,m_1), \dots, (k_q,l_q,m_q),(-k_{q+1},l_{q+1},m_{q+1}), \dots, (-k_n,l_n,m_n)}.
\end{align}
Reciprocally an $n$-particle state in ${\cal H}_{t_1} \otimes {\cal H}_{t_2}^*$ is a linear combination of $n$-particle states in ${\cal H}_{\varrho}$,
\begin{align}
\psi_{(k_1,l_1,m_1), \dots, (k_n,l_n,m_n)} =&
\, (-1)^{n-q} \, (8 \pi)^n  \int \xd \Omega_{k_1} \cdots \int \xd \Omega_{k_n} \, \im^{l_{q+1} + \cdots + l_n- l_1- \cdots -l_q}
\, Y_{l_1}^{m_1}(\Omega_{k_1}) \cdots Y_{l_n}^{m_n}(\Omega_{k_n}) \nonumber\\ 
& \times \psi_{\underline{k}_1, \cdots, \underline{k}_q|\underline{k}_{q+1}, \cdots, \underline{k}_n}.
\end{align}
On the hypercylinder, the incoming or outgoing character of particle states is encoded in the quantum numbers, in particular in the sign of $k$. By means of the isomorphism (\ref{eq:isom}) between the Hilbert states associated with the different geometries considered, a particle in ${\cal H}_{\varrho}$ can be mapped, according to its quantum number, into an incoming or outgoing particle in ${\cal H}_{t_1}$ or ${\cal H}_{t_2}$ respectively.

\section{Summary and outlook}
\label{sec:con}

Let us summarize the results we have obtained. We have applied the GBF of quantum field theory to study a massive scalar field in de Sitter spacetime. Inspired by previous results obtained in Minkowski spacetime \cite{CoOe:spsmatrix,CoOe:smatrixgbf}, two different quantization schemes, associated with different spacetime regions, have been implemented and compared: in the first and more traditional one the region considered is of a time-interval type, namely the boundary of the region is the disjoint union of two equal-time hypersurfaces; in the second scheme, the scalar field is quantized in a region that incorporates key nonstandard features of the GBF, namely the hypercylinder region bounded by one connected and timelike hypersurface. After constructing all the relevant algebraic structures for the free theory, i.e., the Hilbert space, the field propagator and the amplitude associated with the regions in question, we consider the interacting theory, starting with the case of the interaction with a source field and then using functional derivative techniques to treat the general interacting theory. We then show the existence of an isomorphism between the Hilbert spaces defined in the two schemes and provide the explicit correspondence between the multiparticle states defined on the equal-time hypersurfaces with those defined on the hypercylinder.

This work was first motivated by its obvious relevance for the GBF, in order to improve our understanding of its technical and conceptual aspects. Indeed the results presented here constitute the first application of the GBF to a field theory on a curved space. Furthermore, in the particular coordinate systems chosen the de Sitter metric is conformal to the Minkowski metric and we were able to show how all the relevant structures of the GBF in de Sitter space reduce to those computed in Minkowski space in the appropriate asymptotic limit. Such a correspondence not only constitutes a requirement the GBF in de Sitter has to satisfy, but also provides a useful tool to compare the vacuum states defined on the different hypersurfaces considered. What emerges from this is that the field theory of a massive scalar field in the de Sitter metric tends asymptotically to the field theory of a \emph{massless} scalar field in the Minkowski metric. Based on these observations, we can extend to de Sitter space some of the conclusions discussed for the field theory in Minkowski space: In particular the crossing symmetry of the $S$-matrix. As explained in the paper, the asymptotic amplitude for the hypercylinder geometry may be interpreted as the $S$-matrix for the scalar field, and because of the connectedness of the boundary hypersurface no a priori distinction exists between incoming and outgoing states. The usual notion of crossing symmetry becomes consequently implicit in the hypercylinder geometry.

So, the GBF of the field theory in the hypercylinder geometry provides a new perspective on the quantum dynamics in de Sitter space. We have indeed obtained a new representation for the de Sitter invariant Feynman propagator in the hypercylinder region, (\ref{eq:GreenFhyp}), and the corresponding spatially asymptotic boundary condition (\ref{eq:bcGreenFhyp}) for the non-homogeneous Klein-Gordon equation. Moreover, the propagator (\ref{eq:GreenFhyp}) is of the Hadamard form, implying that the vacuum state (\ref{eq:vachyp}) corresponds to the Bunch-Davies vacuum \cite{Bunch:1978yq} defined on the hypercylinder. Hence, apart from its significance for the development of the GBF, our result represents a contribution to the development of the Schr\"odinger representation for quantum fields in curved spacetime. Besides the representation of the Bunch-Davies vacuum, the definition and use of coherent states in the two settings considered here, in the Schr\"odinger representation, not known previously, constitute a novelty of our paper.

As a possible application of the results presented here, we mention the study of the response of the detector moving in the radial direction, i.e. from one hypercylinder to another one of different radius, and the comparison with the well known thermal bath of radiation perceived by an observer moving with proper time $t$, \cite{BiDa:qfcs}.

The implementation of the GBF quantization scheme for more general spacetime regions and boundaries represents a line of future research; in particular compact regions, whose boundary hypersurface includes timelike as well as spacelike parts, will be the main focus. The field theory defined in such kind of regions will provide a fully local description of the quantum dynamics, and it will be interesting the study how a particle concept (associated with the boundary of the region) could be implemented and compare it with the standard notion of particle. A possible direction along this line is the use of a different coordinate system to describe the de Sitter spacetime \cite{BiDa:qfcs}.

We conclude by mentioning a new quantization scheme for the GBF recently proposed by Oeckl in \cite{Oe:newGBF}. The ability to recover the results presented here may constitute an important test for Oeckl's proposal.

\begin{acknowledgments}

The author is grateful to Robert Oeckl for many useful discussions and comments on an earlier draft of this paper. This work was supported in part by CONACyT grant 49093.

\end{acknowledgments}

\appendix

\section{Mode expansion on the hyperceylinder}
\label{sec:apA}

We define in this appendix the inner product for the modes (\ref{eq:uklm}).
Consider the 2-form $\omega$ on the space of solutions of the Klein-Gordon equation in spherical coordinate (\ref{eq:KGhyp}), associated with the hypersurface composed by two hypercylinders of radii $\varrho_1$ and $\varrho_2$, with a bottom cover at $t=t_1$,
\bea
\omega(u_{k,l,m},u_{k',l',m'})&=& \int_{t_1}^{\infty} \xd t \int \xd \Omega \, \sqrt{g^{(3)}_r(t,\Omega)} \, \left( \overline{u_{l,m,n}}(t,r, \Omega) \stackrel{\leftrightarrow}{\partial_r} u_{k',l',m'}(t,r, \Omega)\right)\bigg|^{r=\varrho_2}_{r=\varrho_1} \nonumber\\
&+&  \int_{\varrho_1}^{\varrho_2} \xd r  \, \int \xd \Omega \, \sqrt{g^{(3)}_t(r,\Omega)} \,
 \left( \overline{u_{k,l,m}}(t,r, \Omega) \stackrel{\leftrightarrow}{\partial_t} u_{k',l',m'}(t,r, \Omega) \right)\bigg|_{t=t_1},
\label{eq:ap01}
\eea
where $g^{(3)}_r(t,\Omega)$ and $g^{(3)}_t(r,\Omega)$ are the 3-metrics induced of the hypercylinder of radius $r$ and the disk at time $t$ respectively; also we adopted the following notation,
\be
\left( \overline{u_{k,l,m}}(t,r, \Omega) \stackrel{\leftrightarrow}{\partial_t} u_{k',l',m'}(t,r, \Omega) \right) = \overline{u_{k,l,m}}(t,r, \Omega) \, \frac{\xd}{\xd t} u_{k',l',m'}(t,r, \Omega) - u_{k',l',m'}(t,r, \Omega) \, \frac{\xd}{\xd t} \overline{u_{k,l,m}}(t,r, \Omega).
\ee
The substitution in (\ref{eq:ap01}) the expression (\ref{eq:uklm}) of the modes $u_{k,l,m}$ gives
\bea
\omega(u_{k,l,m},u_{k',l',m'})&=& R^2 \int_{t_1}^{\infty} \xd t \int \xd \Omega \, \overline{Y_l^m(\Omega)} \, Y_{l'}^{m'}(\Omega) \, t \, \overline{H_{\nu}}(k t) \, H_{\nu}(k' t) \, \varrho_2^2 \nonumber\\
&\times & \left( (\overline{c_1} \, j_l(kr) + \overline{c_2} \, n_l(kr)) \stackrel{\leftrightarrow}{\partial_r} (c_1 j_{l'}(k' r) + c_2 n_{l'}(k' r))  \right)\bigg|_{r=\varrho_2} 
\nonumber\\
&-& R^2 \int_{t_1}^{\infty} \xd t \int \xd \Omega \, \overline{Y_l^m(\Omega)} \, Y_{l'}^{m'}(\Omega) \, t \, \overline{H_{\nu}}(k t) \, H_{\nu}(k' t) \, \varrho_1^2  \nonumber\\
&\times & \left( (\overline{c_1} \, j_l(kr) + \overline{c_2} \, n_l(kr)) \stackrel{\leftrightarrow}{\partial_r} (c_1 j_{l'}(k' r) + c_2 n_{l'}(k' r))  \right)\bigg|_{r=\varrho_1}
\nonumber\\
&+& R^2 \int_{\varrho_1}^{\varrho_2} \xd r \, r^2 \, \int \xd \Omega \, \overline{Y_l^m(\Omega)} \, Y_{l'}^{m'}(\Omega) 
(\overline{c_1} \, j_l(kr) + \overline{c_2} \, n_l(kr)) (c_1 j_{l'}(k' r) + c_2 n_{l'}(k' r))
\nonumber\\
&\times & 
t_1 \left( \overline{H_{\nu}}(k t) \stackrel{\leftrightarrow}{\partial_t} H_{\nu}(k' t)\right)\bigg|_{t=t_1}.
\label{eq:ap1}
\eea
The integration in the angular variables gives $\delta_{l,l'} \delta_{m,m'}$. We then notice that 
\begin{multline}
\int_{\varrho_1}^{\varrho_2} \xd r \, r^2 \,(\overline{c_1} \, j_l(kr) + \overline{c_2} \, n_l(kr)) (c_1 j_l(k' r) + c_2 n_l(k' r)) = \\
\frac{r^2}{k^2-k'^2} \left( (\overline{c_1} \, j_l(kr) + \overline{c_2} \, n_l(kr)) \stackrel{\leftrightarrow}{\partial_r} (c_1 j_l(k' r) + c_2 n_l(k' r))  \right)\bigg|_{r=\varrho_1}^{r=\varrho_2},
\end{multline}
where we used the formula 5.11.8 of \cite{Wat:bessel}. We also can rewrite the last line in (\ref{eq:ap1}) as
\bea
t_1 \left( \overline{H_{\nu}}(k t) \stackrel{\leftrightarrow}{\partial_t} H_{\nu}(k' t)\right)\bigg|_{t=t_1} &=&
(k^2-k'^2) \int^{t_1} \xd t \, t\, \overline{H_{\nu}}(k t) \, H_{\nu}(k' t), \nonumber\\
&=& (k^2-k'^2) \, F \left[\,  \overline{H_{\nu}}, H_{\nu}, k,k' \right](t_1),
\eea
where we have introduced the notation
\be
F \left[\,  \overline{H_{\nu}}, H_{\nu}, k,k' \right](t_1) = 
\frac{t_1}{k^2-k'^2} \left( k' \, \overline{H_{\nu}}(k t) \, H_{\nu-1}(k' t) - k \, \overline{H_{\nu-1}}(k t) \, H_{\nu}(k' t) \right).
\ee
The integral of product of Hankel functions can be written as
\be
\int_{t_1}^{\infty} \xd t \, t \, \overline{H_{\nu}}(k t) \, H_{\nu}(k' t) = \frac{2}{\sqrt{kk'}} \delta(k-k') - F \left[\,  \overline{H_{\nu}}, H_{\nu}, k,k' \right](t_1).
\ee
Combining all this in (\ref{eq:ap1}) yields
\bea
\omega(u_{k,l,m},u_{k',l',m'})&=&  \delta_{l,l'} \, \delta_{m,m'} \, \frac{2 R^2 }{\sqrt{kk'}} \, \delta(k-k') \, r^2 \left( (\overline{c_1} \, j_l(kr) + \overline{c_2} \, n_l(kr)) \stackrel{\leftrightarrow}{\partial_r} (c_1 j_l(k' r) + c_2 n_l(k' r))  \right) \bigg|^{r=\varrho_2}_{r=\varrho_1}, \nonumber\\
&=& 0.
\eea
Therefore we can conclude that the following structure
\bea
\omega(u_{k,l,m},u_{k',l',m'}) &=&
R^2 \int \xd \Omega \, \overline{Y_l^m(\Omega)} \, Y_{l'}^{m'}(\Omega) \left( \int_{t_1}^{\infty} \xd t  \, t \, \overline{H_{\nu}}(k t) \, H_{\nu}(k' t) \, r^2 \right. \nonumber\\
&\times&
\left( (\overline{c_1} \, j_l(kr) + \overline{c_2} \, n_l(kr)) \stackrel{\leftrightarrow}{\partial_r} (c_1 j_{l'}(k' r) + c_2 n_{l'}(k' r))  \right)
\nonumber\\
&+&  \int^r \xd r \, r^2 \, (\overline{c_1} \, j_l(kr) + \overline{c_2} \, n_l(kr)) \, (c_1 j_{l'}(k' r) + c_2 n_{l'}(k' r)) \nonumber\\
&\times& \left. t_1 \left( \overline{H_{\nu}}(k t) \stackrel{\leftrightarrow}{\partial_t} H_{\nu}(k' t)\right)\bigg|_{t=t_1} \right),
\eea
expressed in therms of the modes $u_{k,l,m}$ as
\bea
\omega(u_{k,l,m},u_{k',l',m'}) &=& \int_{t_1}^{\infty} \xd t \int \xd \Omega \, \sqrt{g^{(3)}_r(t,\Omega)} \, \left( \overline{u_{l,m,n}}(t,r, \Omega) \stackrel{\leftrightarrow}{\partial_r} u_{k',l',m'}(t,r, \Omega)\right) \nonumber\\
&+&  \int^{r} \xd r  \, \int \xd \Omega \, \sqrt{g^{(3)}_t(r,\Omega)} \,
 \left( \overline{u_{k,l,m}}(t,r, \Omega) \stackrel{\leftrightarrow}{\partial_t} u_{k',l',m'}(t,r, \Omega) \right)\bigg|_{t=t_1},
\eea
is independent of $t_1$ and $r$. The above expression defines the conserved inner product on the space of solutions of the Klein-Gordon equation.

With the result just obtained we can expand the field configuration $\varphi(t,\Omega)$ defined on the hypercylinder according to the formula
\be
\varphi(t,\Omega) = \int_{- \infty}^{\infty} \xd k \sum_{l,m} \, Y_l^m(\Omega) \, |t|^{3/2} \, {\mathscr H}_{\nu}(kt) \, \varphi_{l,m}(k),
\label{eq:a1}
\ee
and the inverse relation for the modes $\varphi_{l,m}$ are expressed as
\be
\varphi_{l,m}(k) =  \int \xd t \, \xd \Omega \, Y_l^{-m}(\Omega) \, |t|^{-1/2} \, \overline{{\mathscr H}_{\nu}(kt)} \, \frac{|k|}{4} \, \varphi(t, \Omega).
\label{eq:a2}
\ee
The appearance of the modulus of $t$ and $k$ in (\ref{eq:a1}) and (\ref{eq:a2}) guarantees the reality of the modes $\varphi(t,\Omega)$.

\section{On the expression of the Feynman propagator in the hypercylinder geometry}
\label{sec:apB}

In this appendix we prove the equivalence of the propagator (\ref{eq:GreenFhyp}) derived in the hypercylinder geometry with the Feynman propagator obtained in the time-interval setting of Section \ref{sec:asymampfree1}. 
The first step consists in summing over $l$ and $m$ in (\ref{eq:GreenFhyp}) according to the following relation (see formulas (B.98) and (B.100) of \cite{Mes:qm1}),
\be
\frac{e^{\im |k| |\underline{x} - \underline{x}'|}}{4 \pi |\underline{x} - \underline{x}'|} = \im |k| \sum_{l,m }Y_l^m(\Omega_z) \overline{Y_l^m(\Omega_{z'})}  \left( \theta(r-r') j_l(|k| r') h_l(|k| r) + \theta(r'-r) j_l(|k| r) h_l(|k| r') \right).
\ee 
So, we obtain for $G_F$ the form
\be
G_F(x,x') = \frac{1}{16 \pi R^2 |\underline{x} - \underline{x}'|} \int_{-\infty}^{\infty} \xd k \, |k| \, e^{\im |k| |\underline{x} - \underline{x}'|} \,  (tt')^{3/2}  {\mathscr H}_{\nu}(k t) \overline{{\mathscr H}_{\nu}(kt') }.
\label{eq:apB1}
\ee
Then, we rewrite the exponential as
\be
e^{\im |k| |\underline{x}-\underline{x}'|} = 
\lim_{\epsilon \rightarrow 0^+} - \frac{\im}{\pi} \int_{-\infty}^{\infty} \xd q \, q \, \frac{e^{\im q |\underline{x}-\underline{x}'| }}{q^2 - k^2 - \im \epsilon}.
\ee
We substitute in (\ref{eq:apB1}), invert the order of the integrals and perform before the integral in $\xd k$,
\bea
G_F(x,x')
&=&   \frac{(tt')^{3/2}}{16 \pi R^2} 
\frac{1}{ |\underline{x}-\underline{x}'|}
\int_{-\infty}^{\infty} \xd q \, q \,e^{\im q |\underline{x}-\underline{x}'| }
\left(
\theta(t'-t) \, \overline{{\mathscr H}_{\nu}}(|q| t') \, {\mathscr H}_{\nu}(|q| t) +
\theta(t-t') \, \overline{{\mathscr H}_{\nu}}(|q| t) \, {\mathscr H}_{\nu}(|q| t') 
\right), \nonumber\\
&=&  \im \frac{(tt')^{3/2}}{8 \pi R^2} 
\int_{0}^{\infty} \xd q \, q \, \frac{\sin \left( q |\underline{x}-\underline{x}'| \right)}{ |\underline{x}-\underline{x}'|} 
\left(
\theta(t'-t) \, \overline{{\mathscr H}_{\nu}}(q t') \, {\mathscr H}_{\nu}(q t) +
\theta(t-t') \, \overline{{\mathscr H}_{\nu}}(q t) \, {\mathscr H}_{\nu}(q t') 
\right), \nonumber\\
\eea
which coincides with (\ref{eq:GF1}) since $\overline{{\mathscr H}_{\nu}}(x) \, {\mathscr H}_{\nu}(y) = \overline{H_{\nu}}(x) \, H_{\nu}(y)$ due to the relation (\ref{eq:modHankel}). This concludes the proof.

\bibliographystyle
{amsordx}
\bibliography{stdrefs,refs2}

\end{document}